\documentclass[fleqn,usenatbib]{mnras}

\usepackage{newtxtext,newtxmath}
\usepackage[T1]{fontenc}
\usepackage{comment}

\DeclareRobustCommand{\VAN}[3]{#2}
\let\VANthebibliography\thebibliography
\def\thebibliography{\DeclareRobustCommand{\VAN}[3]{##3}\VANthebibliography}

\usepackage{graphicx}	% Including figure files
\usepackage{amsmath}	% Advanced maths commands
\usepackage{physics}
\usepackage{float}
\usepackage{hyperref}
\usepackage{multirow, tabularx}
\usepackage{array}
\usepackage[table, svgnames, dvipsnames]{xcolor}
\usepackage{makecell, cellspace, caption}
\usepackage{mathtools}
\usepackage{url}
\usepackage{braket}
\usepackage{threeparttable}
\usepackage{orcidlink}
\usepackage{bm}         % For bold math symbols 
\usepackage[normalem]{ulem} % normalem: keep \emph italic (ulem only provides \sout)

\usepackage{zebra-goodies}              % Mark comments/notes. 
\zebranewnote{question}{orange}
\zebranewnote{revised}{red}
\newcommand\mpc{\, h^{-1}{\rm {Mpc}}}

\newcommand\kms{\,{\rm {km\, s^{-1}}}}

\newcommand\msun{\, h^{-1}{\rm M_\odot}}

\newcommand\softwarestyle{\sc}

\newcommand\textem{\em}                 % emphasis in text
\newcommand\bfem{\bf\em}                % emphasis in text, bold
\newcommand\figem{\bf}                  % emphasis in fig caption
\newcommand\software{\sc}
\title[Unbound matter mediates halo secondary bias]{Unbound matter in the circum-halo medium: a common mediator of halo secondary bias}

\author[Ma et al.]{
    Qinglin Ma\orcidlink{0009-0000-5148-9457} $^{1}$\thanks{E-mail: maqlin2023@gmail.com},
    Yangyao Chen\orcidlink{0000-0002-4597-5798} $^{2,3}$\thanks{E-mail: yangyaochen.astro@foxmail.com} 
    and Houjun Mo\orcidlink{0000-0001-5356-2419}$^{4}$
\\
$^{1}$Department of Astronomy, Tsinghua University, Beijing 100084, China\\
$^{2}$School of Astronomy and Space Science, Nanjing University, Nanjing, Jiangsu 210093, China\\
$^{3}$Key Laboratory of Modern Astronomy and Astrophysics, Nanjing University, Ministry of Education, Nanjing, Jiangsu 210093, China\\ 
$^{4}$Department of Astronomy, University of Massachusetts, Amherst, MA 01003, USA
}

\date{Accepted XXX. Received YYY; in original form ZZZ}

\pubyear{\the\year{}}

\begin{document}
\label{firstpage}
\pagerange{\pageref{firstpage}--\pageref{lastpage}}
\maketitle

% Use macros, \msun and \Msun, whenever possible.
\begin{abstract}
Halo secondary bias describes the dependence of halo clustering on properties
beyond mass. Here we report that a new environmental metric, the unbound matter
fraction ($f_{\rm un}$) in the $2$--$3\,R_{\rm vir}$ shell around a halo, can
largely account for secondary bias. 
Using haloes with $M_{\rm vir}=10^{10.5}$--$10^{14.5}\msun$ at $z=0$ obtained 
from TNG300-1-Dark, we find that controlling for $f_{\rm un}$ reduces the 
secondary bias of formation time, concentration, spin, shape and velocity 
anisotropy to a residual correlation of $\lesssim 4\%$, outperforming 
other environmental metrics such as the tidal anisotropy for
spin and velocity anisotropy.
Applying a stacking method to the circum-halo medium (CHM), we identify two 
distinct flow patterns: low-$f_{\rm un}$ haloes co-move with coherent flows, 
with the CHM converging onto the haloes; in contrast, high-$f_{\rm un}$ haloes 
move through dynamically hot, partially virialized structures, with the CHM passing 
by rather than being accreted. This dichotomy points to two distinct regimes of
secondary bias, a `cold' regime and a `hot' regime, in which tidal and inertial
forces act as the main drivers, respectively, with the dominant mechanism
migrating from the former to the latter as haloes assemble into massive structures.
At fixed large-scale environment, $f_{\rm un}$ further indicates the presence 
of nearby neighbours and traces competitive accretion.
Our results highlight the role of unbound matter as a mediator of environmental 
effects on haloes, and point to the physical mechanisms through which these 
effects are mediated.
\end{abstract} 

% Select between one and six entries from the list of approved keywords.
% Don't make up new ones.

%%%%%%%%%%%%%%%%%%%%%%%%%%%%%%%%%%%%%%%%%%%%%%%%%%

%%%%%%%%%%%%%%%%% BODY OF PAPER %%%%%%%%%%%%%%%%%%

\section{Introduction} \label{sec:introduction}

In the $\Lambda$-cold-dark matter ($\Lambda$CDM) cosmology,
dark matter haloes are the fundamental building blocks of cosmic structure and 
provide the sites where galaxies form and evolve \citep[e.g.][]{whiteCoreCondensationHeavy1978,moGalaxyFormationEvolution2010}.
Understanding the clustering of dark matter haloes is therefore central to both cosmology 
and galaxy formation.
The dependence of halo bias on halo mass itself has a long theoretical history: 
it was first derived from the extended Press-Schechter formalism by 
\citet{moAnalyticModelSpatial1996} and tested extensively against simulations, 
and was later refined by the ellipsoidal-collapse calculation of \citet{shethEllipsoidalCollapseImproved2001}. 
Motivated by this success, most empirical descriptions of the galaxy-halo connection, 
including Halo Occupation Distribution (HOD) and Conditional Luminosity Function 
(CLF) models, assume that halo clustering is determined solely by halo mass 
\citep{berlindHaloOccupationDistribution2002,yangConstrainingGalaxyFormation2003}.
This assumption underpins many statistical approaches for interpreting galaxy 
surveys and connecting observed galaxies to the underlying dark matter distribution, 
making its validity central to modern models of the galaxy-halo connection 
\citep[e.g.][]{wechslerConnectionGalaxiesTheir2018}.

Yet mass may not be the whole story: the ellipsoidal-collapse calculation already 
implied a secondary dependence of halo clustering on formation time through the 
local tidal field, as pointed out explicitly by \citet{Sheth2004OnEnvironmentalDependence}, 
before such a dependence was discovered in simulations. 
This mass-only paradigm was eventually challenged by the discovery of halo assembly 
bias in N-body simulations, whereby haloes of the same mass exhibit different 
clustering amplitudes depending on their assembly histories \citep{gaoAgeDependenceHalo2005}. 
Subsequently, assembly bias has been identified for an increasingly diverse set 
of halo properties, including concentration, spin, shape, velocity anisotropy, subhalo 
occupation, and other structural and dynamical indicators
\citep[e.g.][]{wechslerDependenceHaloClustering2006,gaoAssemblyBiasClustering2007,
jingDependenceDarkHalo2007,wangEnvironmentalDependenceCold2007,liHaloFormationTimes2008,
Faltenbacher2010AssemblyBias,Lazeyras2017LargescaleAssemblyBias,
maoAssemblyBiasExploring2018,xuDependenceHaloBias2018,sato-politoDependenceHaloBias2019,
hanMultidimensionalDependenceHalo2019,Johnson2019SecondarySpinBias,
chenRelatingStructureDark2020,Wang2021EvaluatingOrigins,
BalagueraAntolinez2024SecondaryHaloBias,Ramakrishnan2025MultidimensionalHalo,
Wang2026HaloAssemblyBias,WangAlternativeassemblybias2026}. 
Since this phenomenon is not solely caused by the halo assembly history, it is also 
broadly referred to as secondary bias.
The growing list of halo properties that exhibit secondary bias 
suggests that halo clustering is
sensitive to a much richer set of halo properties than halo mass alone.

Although the existence of assembly bias is firmly established, its physical origin 
remains an open question. 
Early studies mainly sought to explain individual forms of secondary bias through 
distinct physical mechanisms, such as peak theory 
\citep{Zentner2007TheExcursionSet,dalalHaloAssemblyBias2008}, splashback objects 
\citep[e.g.][]{wangDistributionEjectedSubhalos2009,Ludlow2009TheUnorthodoxOrbits,
sunayamaScaledependenceHaloAssembly2016,villarrealImmitigableNatureAssembly2017,
mansfieldThreeCausesLowmass2020,tucciPhysicalOriginsLowmass2021}, 
tidal suppression \citep[e.g.][]{hahnTidalEffectsEnvironment2009,
wangInternalPropertiesEnvironments2011,Shi2015FlowPatternsaround,
salcedoSpatialClusteringDark2018,mansfieldThreeCausesLowmass2020,
Smith2024ReversingArrestedDevelopment}, protohaloes \citep[e.g.][]{Lee2024DependenceHaloBias,MonteroDorta2025ExploringPhysicalOrigins,Nikakhtar2025ProtohalosConnection},
and cosmic web \citep[e.g.][]{Shen2006ExcursionSetModel,Yang2017RevealingCosmicWebdependentHaloBias,
Borzyszkowski2017ZOMGHowCosmic,Musso2018HowDoesCosmic,MonteroDorta2024TheDependenceAssembly,ColomaNadal2024HierarchicalCosmicWeb};
see also \citet{buschTessellationLevelTreeCharacterizingNested2020} 
for a direct characterization of the nested hierarchy of density peaks. 
These mechanisms do not constitute a unified theoretical framework for halo 
assembly bias; rather, they imply that different forms of secondary bias may 
originate from distinct, yet closely related, physical processes operating 
between haloes and their surrounding large-scale environments \citep{salcedoSpatialClusteringDark2018,mansfieldThreeCausesLowmass2020}. 

Several studies have sought a unified framework rather than separate 
explanations for individual halo properties. An early step in this direction 
was taken by \citet{wangInternalPropertiesEnvironments2011}, who proposed that 
the local tidal field regulates halo mass accretion and thereby simultaneously 
influences multiple secondary halo properties \citep[see also][]{hahnTidalEffectsEnvironment2009}. 
In this framework, the correlation between halo internal properties and 
large-scale clustering stems from the strength of the local tidal field \citep{wangEnvironmentalDependenceCold2007,Shi2015FlowPatternsaround}.  
Building upon this idea, \citet{paranjapeHaloAssemblyBias2018} introduced tidal 
anisotropy as a more fundamental environmental variable
\citep[see also][]{ramakrishnanCosmicWebAnisotropy2019,Ramakrishnan2020SeparateUniversecalibration}. They showed that the anisotropy 
of the surrounding tidal field accounts for much of the dependence of halo clustering 
on a broad range of secondary halo properties, suggesting that many manifestations 
of secondary bias are driven by the interaction between haloes and the anisotropic 
cosmic web \citep{Borzyszkowski2017ZOMGHowCosmic,Musso2018HowDoesCosmic}.

% Distinguish unbound particles from unbound matter -- the latter is more 
% general, while the former involves numerical representation.
Although the tidal field and its anisotropy provide a successful environmental 
description of halo secondary bias, they are external quantities that characterize 
the large-scale environment surrounding a halo, and do not directly record 
how environmental forces modify halo properties. 
A promising clue was recently provided by
\citet{MaInterpretingStrongClustering2026}, who showed that the spin bias of 
low-mass haloes changes qualitatively depending on whether unbound matter 
in haloes is included in the halo definition. Besides spin,
unbound matter also affects the concentration bias, but has little effect on
the shape bias \citep{MaAnisotropicSecondaryBias2026}.
These findings suggest that unbound matter may provide
the internal link missing from the purely external tidal description,
i.e. a quantity that records how the environment acts on haloes.
Whether unbound matter merely modifies individual bias measurements or traces a
physical process that generates the secondary bias as a whole has not
been addressed.

% Section XXX is formally used by MNRAS, but we have to keep a consistency
% with the rest of the paper. 
% Keep this outline short -- do not describe subsections.
In this paper, we explore the role of the unbound mass fraction, $f_{\rm un}$, 
in halo secondary bias using the {\software TNG300-1-Dark} simulation.
\S\ref{sec:simulation-data} describes the simulation and the measurement of 
halo properties.
\S\ref{sec:secondary_bias} shows that controlling for $f_{\rm un}$ largely 
removes the secondary bias associated with the internal halo properties considered, 
and benchmarks $f_{\rm un}$ against the two leading environmental indicators 
proposed in previous studies. 
\S\ref{sec:physics-of-fun-vs-b} identifies the
circum-halo flow patterns traced by $f_{\rm un}$ and proposes the physical
mechanisms that shape them.
\S\ref{sec:summary} summarizes our results and discusses
implications for the galaxy--halo connection and observational tests.

\section{Simulation and halo properties}
\label{sec:simulation-data}

\subsection{The simulation and halo sample}

Throughout this paper,
we use the {\software TNG300-1-Dark} simulation, carried out as part 
of the IllustrisTNG project \citep[][]{springelFirstResultsIllustrisTNG2018,pillepichFirstResultsTNG502019,
nelsonFirstResultsIllustrisTNG2018,nelsonIllustrisTNGSimulationsPublic2019,marinacciFirstResultsIllustrisTNG2018,naimanFirstResultsIllustrisTNG2018}. 
The simulation evolves the density field 
using $2500^{3}$ dark-matter particles, each with a mass of $7\times 10^{7} \mathrm{M}_{\odot}$,
in a periodic box with a side length of $205 \mpc$. The simulation adopts the Planck-15 cosmology \citep{Planck2016CosmologicalParameters}, with $\Omega_m$ = 0.3089, $\Omega_b$ = 0.0486, $\Omega_{\Lambda}$ = 0.6911, $H_0 = 100\,h\kms\,{\rm Mpc}^{-1}$ with $h = 0.6774$, $\sigma_8 = 0.8159$ and $n_s = 0.9667$.
Haloes are identified using the {\software FoF} algorithm with a scaled linking length of 0.2 \citep{davisEvolutionLargescaleStructure1985},
and subhaloes are identified using the {\software Subfind} algorithm \citep{springelPopulatingClusterGalaxies2001}. 
Subhalo merger trees are constructed using the {\software SubLink} algorithm \citep{Rodriguez-Gomez2015MergerRateGalaxies}.
The centre of a halo is defined as the position of the most bound particle 
in the central subhalo identified by {\software Subfind}. 
The virial radius ($R_{\rm vir}$) of a halo is defined as the radius within 
which the enclosed matter density equals the virial density threshold 
given by the spherical collapse model \citep{bryanStatisticalPropertiesXRay1998},
and the virial mass ($M_{\rm vir}$) of the halo is the total mass enclosed within 
$R_{\rm vir}$.
The virial velocity $V_{\rm vir}$ is defined as 
$V_{\rm vir}=\sqrt{GM_{\rm vir}/R_{\rm vir}}$. 
% Do not use term like `halo velocity', since it is confusing with the motional 
% velocity of the halo (a vector).
Our sample consists of distinct haloes with $M_{\rm vir} = 10^{10.5}$--$10^{14.5}\msun$ 
identified at $z=0$. The lower mass limit ensures that 
haloes are resolved with $\gtrsim 700$ particles, sufficient for measuring 
the properties described below.

\subsection{Secondary properties of haloes}
\label{ssec:halo_property}

We define the following internal properties of each halo, which characterize
its assembly and structure, and have been found to correlate with halo
clustering:
\begin{itemize}
\item The half-mass formation redshift, $z_{1/2}$ (hereafter also referred to as the formation time), is the redshift at which a halo first reaches half of its final virial mass at $z=0$. We determine this 
redshift by linearly interpolating the mass-redshift relation between two 
adjacent snapshots in which the halo mass crosses 
$M_{\rm vir}/2$ \citep{gaoAgeDependenceHalo2005}.

\item The concentration, $c_{\rm vir}$, is computed from the virial velocity, $V_{\rm vir}$, and the maximum circular velocity, $V_{\rm max}$, both measured using all particles, through the relation
\begin{equation}
    \frac{c_{\rm vir}}{f(c_{\rm vir})} = \frac{V_{\rm max}^2}{V_{\rm vir}^2} \frac{R_{\rm max}/R_{\rm s}}{f(R_{\rm max}/R_{\rm s})},
\end{equation}
where $f(x) \equiv \ln (1+x) - x/(1+x)$, $R_{\rm max}$ is the radius at which $V_{\rm max}$ is reached and $R_{\rm s} \equiv R_{\rm vir}/c_{\rm vir}$ is the scale radius. This concentration estimator is robust to variations in the detailed halo profile \citep{klypinDarkMatterHalos2011}. 
    %Therefore, it is more robust than the concentration derived from the NFW profile \citep{navarroStructureColdDark1996,navarroUniversalDensityProfile1997}. 

\item The spin parameter, $\lambda$, is a dimensionless measure of the total angular momentum \citep{bullockProfilesDarkHaloes2001}, defined as
    \begin{equation}
\lambda=\frac{|J|}{\sqrt{2} M_{\mathrm{vir}} V_{\mathrm{vir}} R_{\mathrm{vir}}},
\end{equation}
where $|J|$ is the magnitude of the total angular momentum of the particles within $R_{\rm vir}$. 
Spin bias is sensitive to the definition of the spin parameter, particularly to whether 
unbound particles are included \citep{MaInterpretingStrongClustering2026}. We therefore calculate the spin parameter using 
two particle sets: bound particles ($\lambda_{\rm b}$) and all (bound plus unbound) particles 
($\lambda_{\rm a}$) within $R_{\rm vir}$. To determine whether a particle is bound
to a halo, we use all particles within $4\,R_{\rm vir}$ of its centre 
to compute each particle's specific kinetic energy (including the Hubble flow) and specific potential 
energy (see \citealt{MaInterpretingStrongClustering2026} for details), with the 
zero point of the potential energy set at the most distant particle among
those within $4\,R_{\rm vir}$. Tests with alternative choices (e.g. using
an aperture of $5\,R_{\rm vir}$) yield similar results.
A particle is classified as unbound if its total specific energy (kinetic plus potential) 
is positive, and as bound otherwise. Note that particles beyond $R_{\rm vir}$ 
can be classified as gravitationally bound and may therefore be accreted into 
the halo in the future. We use the same classification
when computing the halo properties defined below.

\item Shape parameters are obtained from the weighted inertia tensor,
    \begin{equation}
    \tilde{I}_{ij} \equiv \sum_{k} \frac{x_{i, k} x_{j, k}}{r_k^2},
    \label{eq:shape}
\end{equation}
computed iteratively \citep{Allgood2006ShapeDarkMatter} 
using all bound particles enclosed within $R_{\rm vir}$.
Here, $x_{i,k}$ is the displacement of the $k$-th particle from the halo centre 
along the $i$-th spatial dimension in the simulation frame;
% we switch the notation to a_1, a_2 and a_3, since b is used for halo bias
the length of the $i$-th ellipsoidal axis, $a_i$, is proportional to the square root of the 
$i$-th eigenvalue of $\tilde{I}_{ij}$, with the axes ordered such that 
$a_1 \geq a_2 \geq a_3$;
% similarly, we do not use axis ratios $q\equiv b/a$ and $s\equiv c/a$,
% since Q is used to denote outer c/a (confusing with q = b/a).
the weighting uses the ellipsoidal distance $r_k$, defined as
\begin{equation}
    r_k= a_1 \sqrt{\sum_{i=1}^{3} \left(\frac{x'_{i, k}}{a_i}\right)^2}\,,
\end{equation}
where $x'_{i,k}$ is the displacement of the $k$-th particle from the halo centre along the 
$i$-th eigenvector of $\tilde{I}_{ij}$. Following common practice, we use the minor-to-major 
axis ratio, $q = a_3/a_1$, 
to characterize the departure of the halo from sphericity, 
with small $q$ indicating an elongated shape and 
$q = 1$ indicating a spherical shape.

\item The velocity anisotropy is defined as $\beta=1-\sigma_t^2 / 2 \sigma_r^2$ \citep{Binney1987GalacticDynamics}, 
where $\sigma_t$ and $\sigma_r$ are the tangential and radial velocity dispersions,
respectively.
We calculate these dispersions by projecting the velocity of 
each particle
along and perpendicular to the radial direction and then computing the standard deviation of each component over all 
bound particles within $R_{\rm vir}$. 
A value of $\beta=0$ indicates an isotropic velocity distribution, while $\beta>0$ indicates radially biased orbits and $\beta<0$ indicates tangentially biased orbits. 
\end{itemize}

We use the following properties to quantify halo clustering
and characterize the surrounding environment, providing diagnostics
of the origin of secondary bias:
\begin{itemize}

\item The halo bias, $b$, measured in a halocentric spherical shell spanning
the radial range $[r_{\rm min}, r_{\rm max}]$ 
is defined as the ratio of the halo-matter cross-correlation function 
$\xi_{\rm hm}$ to the matter auto-correlation function $\xi_{\rm mm}$
within the shell:
\begin{equation}
    b = \frac{\xi_{\rm hm}}{\xi_{\rm mm}} 
        = \frac{\delta}{\xi_{\rm mm}}, 
\label{eq:bias}
\end{equation}
where $\delta = \rho /\bar\rho-1$ is the matter overdensity,
estimated using the particles within the shell, and $\bar\rho$ is the mean matter 
density of the Universe. 
In the main text, we focus on halo bias over the radial range 
$5$--$12 \mpc$, a quasi-linear regime \citep[e.g.][]{gaoAgeDependenceHalo2005} 
that has been widely used in the literature to measure halo bias
\citep[e.g.][]{chenRelatingStructureDark2020,zhangUnexpectedClusteringPattern2025}.
In Appendix~\ref{app:halo-boundary}, we also examine the bias profile,
$b(r)$, obtained using shells with finite thickness 
spanning $[r-\Delta r/2, r+\Delta r/2]$.
% The bias factor $b$ used for the Spearman correlation analyses below is obtained by averaging the individual bias profiles over $r = 5$--$12 \mpc $, where the bias is approximately scale-independent.  
% We also use the average halo bias
% \begin{equation}
% \langle b(r)\rangle = \langle\delta(r)\rangle/\xi_{\rm mm}(r),
% \label{eq:average_bias}
% \end{equation}
% obtained in two steps: haloes are first binned by a property and averaged within each bin, and % the result is then averaged over the radial range $r = 5$--$12 \mpc $.

\item The tidal anisotropy factor, $\alpha(\bm{x})$, at a spatial location $\bm{x}$ 
quantifies the asymmetry of the gravitational environment at that location
\citep{paranjapeHaloAssemblyBias2018}. 
We follow previous studies \citep{hahnPropertiesDarkMatter2007,ForeroRomero2014CosmicWebAlignments}
and compute $\alpha(\bm{x})$ from the deformation tensor, $\tilde{T}_{ij}(\bm{x})$, defined 
as the Hessian of the gravitational potential $\phi$ 
(scaled by a factor of $4\pi G\bar\rho$): 
\begin{equation}
\tilde{T}_{ij}(\bm{x})=\frac{\partial^2\phi(\bm{x})}{\partial x_i\partial x_j} 
\,,
\label{eq:tidal-tensor-def}
\end{equation}
where $\phi$ is obtained by solving the Poisson equation $\nabla^2\phi(\bm{x})=\delta(\bm{x})$ 
using the overdensity field $\delta(\bm{x})$ derived from all particles in 
the simulation. Here, $x_i$ and $x_j$ are the Cartesian coordinates along
the $i$-th and $j$-th dimensions, respectively, and
$G$ is the gravitational constant.
Since haloes of different sizes are affected by tidal fields on different 
scales, we compute $\tilde{T}_{ij}$ for each halo at a smoothing scale proportional to 
its virial radius. We smooth $\delta(\bm{x})$ using a Gaussian window with 
a kernel size of $4\,R_{\rm vir}$, a choice found to yield
the strongest signal of secondary bias 
\citep{paranjapeHaloAssemblyBias2018}.
We then use the smoothed overdensity field to compute $\phi(\bm{x})$ and 
$\tilde{T}_{ij}(\bm{x})$, evaluating the tensor at the halo centre
to characterize the tidal environment of the halo.
We diagonalize $\tilde{T}_{ij}$ to obtain the eigenvalues
$\lambda_1\geq\lambda_2\geq\lambda_3$ and compute the tidal anisotropy 
factor as
\begin{equation}
\alpha= \frac{\sqrt{g^2}}{1+\delta}\,,
\end{equation}
where $g^2= \left[(\lambda_1-\lambda_2)^2+(\lambda_2-\lambda_3)^2+(\lambda_1-\lambda_3)^2\right]/2$ 
is the tidal shear, and $\delta$ is the overdensity at the location of the halo.
% $\delta=\lambda_1+\lambda_2+\lambda_3$ is the overdensity.

\item The environmental axis ratio, $q_{\rm e} \equiv a_3/a_1$, is measured in the same way as 
the halo axis ratio ($q$), but using all particles at halocentric distances 
in the range $r = 2$--$4\,R_{\rm vir}$ \citep{MaAnisotropicSecondaryBias2026}. 
We omit the iterative procedure, setting 
$r_k = 1$ for all particles in the computation of the inertia tensor (Eq.~\ref{eq:shape}) 
and evaluating the ellipsoidal axes directly from that tensor. Our tests show 
that including the iteration yields similar values of $q_{\rm e}$.
\end{itemize}

\begin{figure}
    \includegraphics[width=\columnwidth]{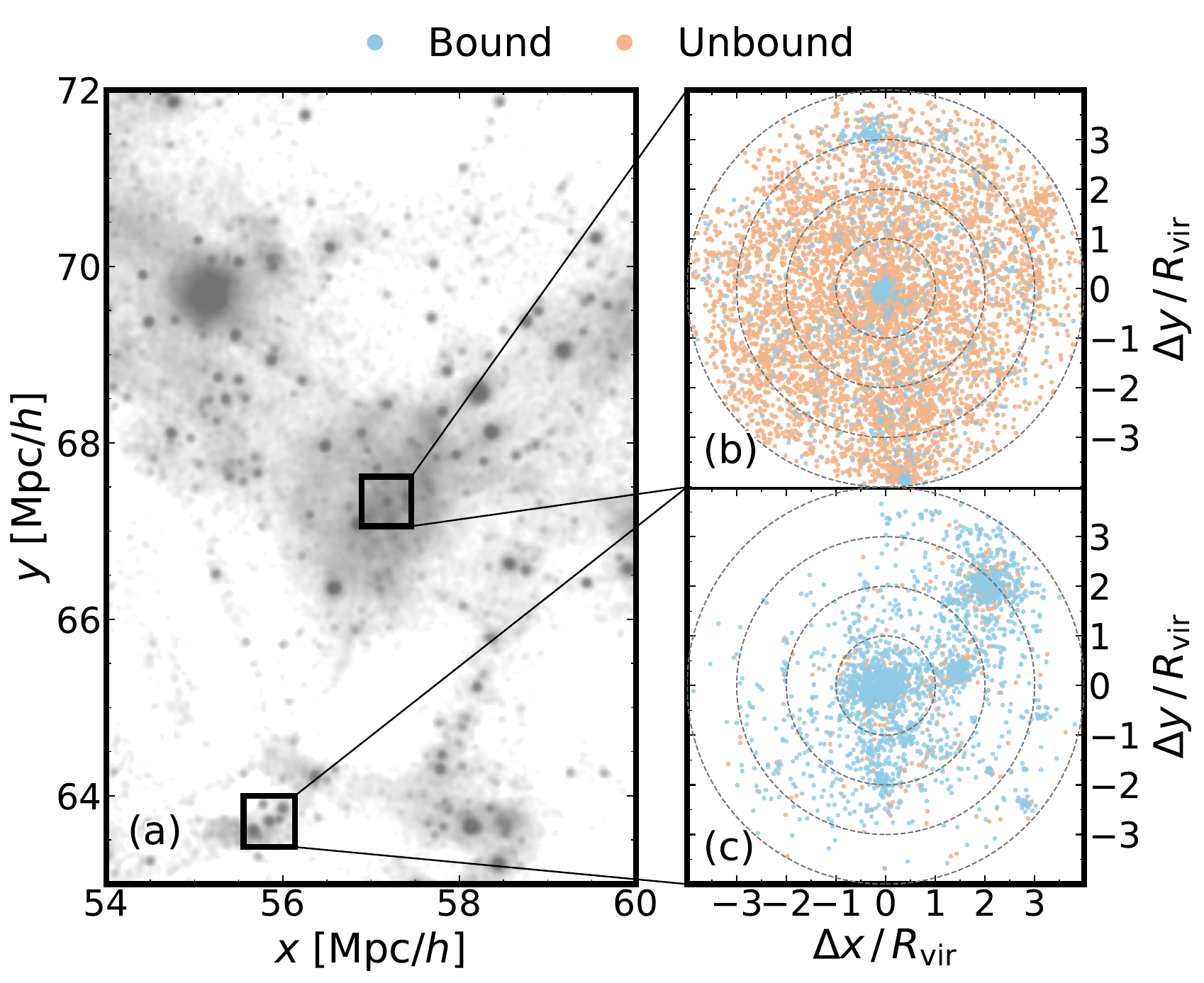}
    \caption{
    {\figem Two example haloes with distinct unbound mass fractions.}
    {\figem (a)}, distribution of dark matter in a slice of the simulation box
    extending $\pm 0.3\mpc$ from the midplane.
    Here the grey background shows the
    % a scalar cannot be projected
    dark matter surface density
    in this slice. 
    {\figem (b)}, {\figem (c)}, zoom-in views of the two example haloes marked by black frames in panel {\figem (a)}.
    The two haloes have comparable masses ($M_{\rm vir} \approx 10^{10.6} \msun$), 
    but contrasting unbound mass fractions of $f_{\rm un}=0.87$ and 
    $0.21$, respectively, measured within $4\,R_{\rm vir}$.
    Each zoom-in panel is centred on the halo centre, 
    with four dashed circles marking the halo-centric distances 
    of $1$, $2$, $3$ and $4\,R_{\rm vir}$, respectively;
    individual bound and unbound particles are represented by blue and orange
    dots, respectively. This figure
    illustrates the contrast in unbound mass fraction between the two haloes,
    and the different large-scale environments they reside in (see \S\ref{data:unbound_particle_fraction} for details).
    }
    \label{fig:fun_example}
\end{figure}

\begin{figure}
    \includegraphics[width=\columnwidth]{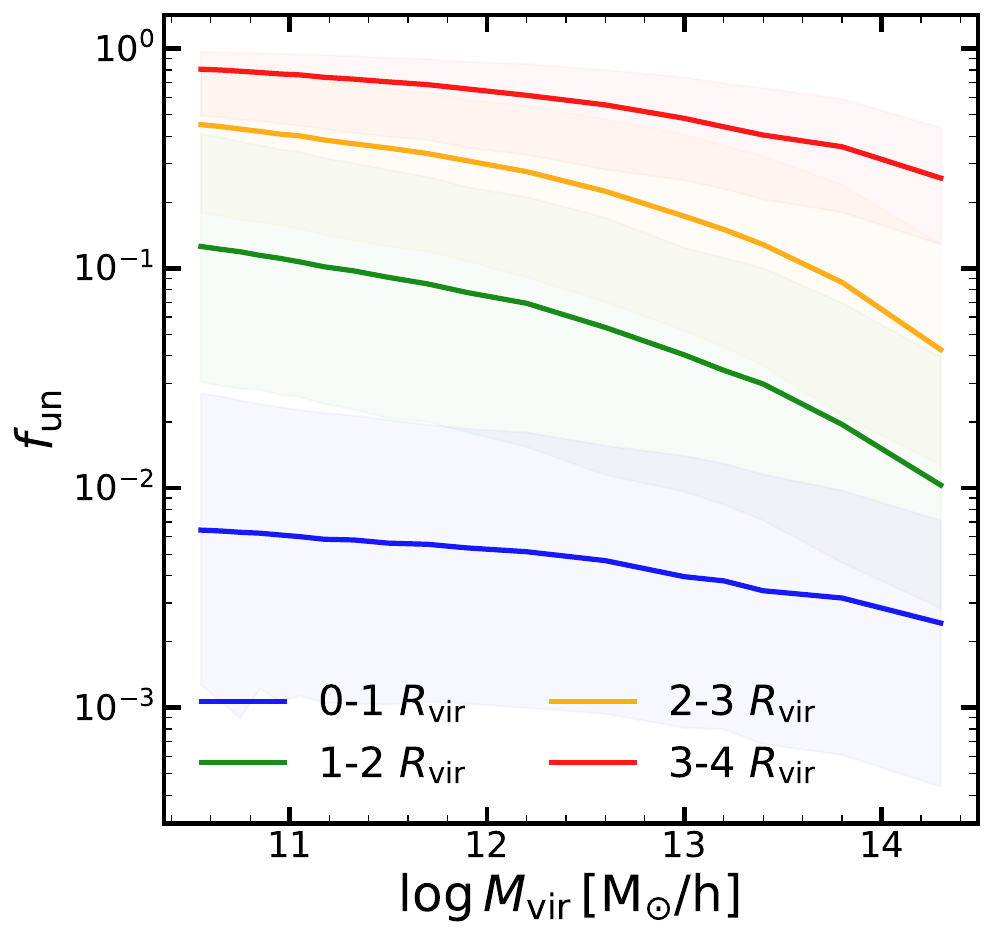}
    \caption{
    {\figem Fraction of unbound mass in and around haloes.}
    Here we show the unbound mass fraction
    ($f_{\rm un}$) as a function of halo mass ($M_{\rm vir}$) measured 
    within shells spanning different halo-centric distances:
    $0$--$1\,R_{\rm vir}$ (blue), 
    $1$--$2\,R_{\rm vir}$ (green), $2$--$3\,R_{\rm vir}$ (orange)
    and $3$--$4\,R_{\rm vir}$ (red). Solid lines show the median values and the 
    shaded regions the $25^{\rm th}$--$75^{\rm th}$ percentiles.
    This figure shows that $f_{\rm un}$ is higher for lower-mass 
    haloes and increases with distance from the halo centre
    (see \S\ref{data:unbound_particle_fraction} for details).
    }
    \label{fig:fun_props}
\end{figure}

\subsection{The unbound mass fraction}
\label{data:unbound_particle_fraction}

The unbound mass fraction, $f_{\rm un}$, in a spherical shell centred on a halo 
is defined as the total mass of unbound particles divided by the total mass of 
all particles within that shell. To illustrate the definition, 
Fig.~\ref{fig:fun_example}(a) presents two example haloes with comparable 
masses ($M_{\rm vir} \approx 10^{10.6} \msun$) but contrasting $f_{\rm un}$ 
(measured within $4\,R_{\rm vir}$). Panel (a) shows their large-scale 
environments: the low-$f_{\rm un}$ halo resides in a void, whereas the 
high-$f_{\rm un}$ halo lies near a node, at the root of a filament. 
Panels (b) and (c) zoom in on the high- and low-$f_{\rm un}$ haloes, respectively:
both haloes have a significant number of bound particles within $R_{\rm vir}$;
around the high-$f_{\rm un}$ halo, the projected matter density is notably 
higher and dominated by unbound particles, whereas the low-$f_{\rm un}$ halo is embedded in a sparser field consisting mainly of bound particles. 
The contrast between the bound-dominated interior and the unbound-rich outskirts motivates examining $f_{\rm un}$ in radial shells. 
To examine the radial dependence,
we calculate $f_{\rm un}$ for each halo in shells spanning four radial ranges:
$(0$--$1)\,R_{\rm vir}$, $(1$--$2)\,R_{\rm vir}$, $(2$--$3)\,R_{\rm vir}$ and $
(3$--$4)\,R_{\rm vir}$.
Fig.~\ref{fig:fun_props} shows the median $f_{\rm un}$ as a function of halo mass. 
The unbound mass fraction is higher for lower-mass haloes and increases with distance from the halo centre. 
Within the virial radius, the median $f_{\rm un}$ is less than $0.008$, 
consistent with results in the literature
\citep[see fig.~3 in][]{behrooziUnboundParticlesDark2013}. 
In the outermost shell, $(3$--$4)\,R_{\rm vir}$, $f_{\rm un}$
reaches $0.8$ for low-mass haloes but only about $0.3$ for 
high-mass haloes, a mass dependence reflecting the stronger gravity of more 
massive haloes in competing against external forces, as discussed in 
\S\ref{sec:physics-of-fun-vs-b}.

As we will see, $f_{\rm un}$ encodes most of the secondary bias, while other 
assembly and structural properties produce little additional bias once 
$f_{\rm un}$ is taken into account. This points to $f_{\rm un}$ as a
mediator through which environmental effects propagate into haloes, and
thereby as a route to the origin of secondary bias.

\begin{figure*}
    \includegraphics[width=\textwidth]{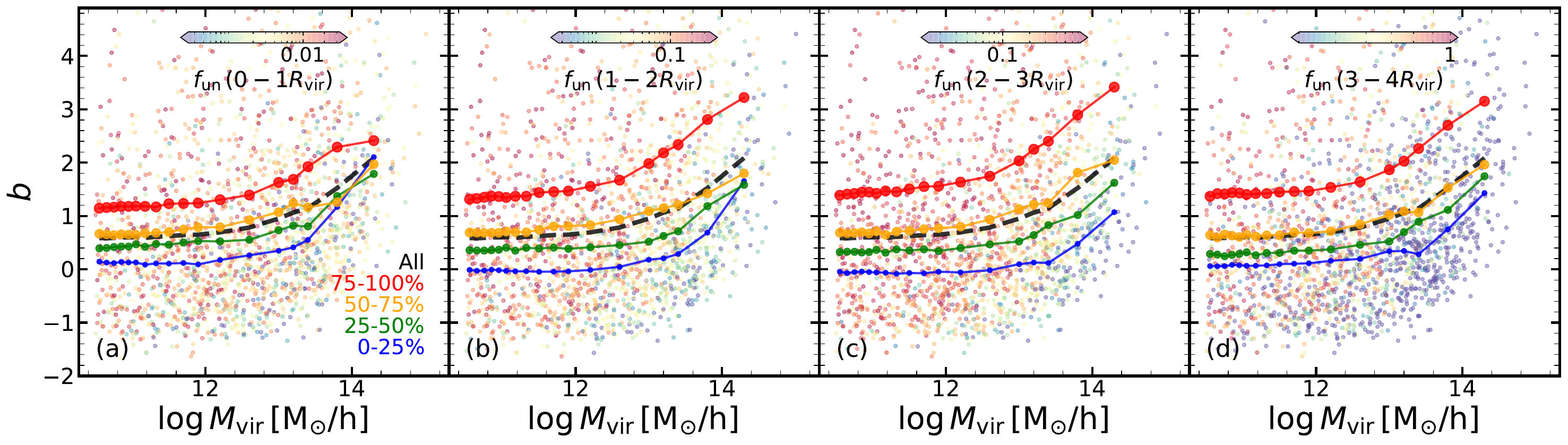}
    \caption{
    {\figem Relation between unbound mass fraction $f_{\rm un}$ and halo bias $b$.}
    {\figem (a)}--{\figem (d)} show the results obtained using $f_{\rm un}$ measured 
    in shells spanning different halo-centric distances.
    In each panel, individual dots in the background represent randomly selected haloes, 
    coloured by the $f_{\rm un}$ value measured in the corresponding 
    shell, according to the colour bar on the top.
    Curves show the average halo bias $b$ of haloes in bins 
    of $M_{\rm vir}$, separately for haloes in the four quartiles of 
    $f_{\rm un}$ in each mass bin (coloured solid) and for all haloes 
    in that bin (black dashed).
    This figure shows that haloes with higher $f_{\rm un}$ are more 
    strongly clustered, and that such a bias-$f_{\rm un}$ correlation 
    persists in all radial ranges and halo masses examined 
    (see \S\ref{ssec:fun-bias} for details).
    }
    \label{fig:diffm_un_bias}
\end{figure*}

\begin{figure*}
    \includegraphics[width=0.925\textwidth]{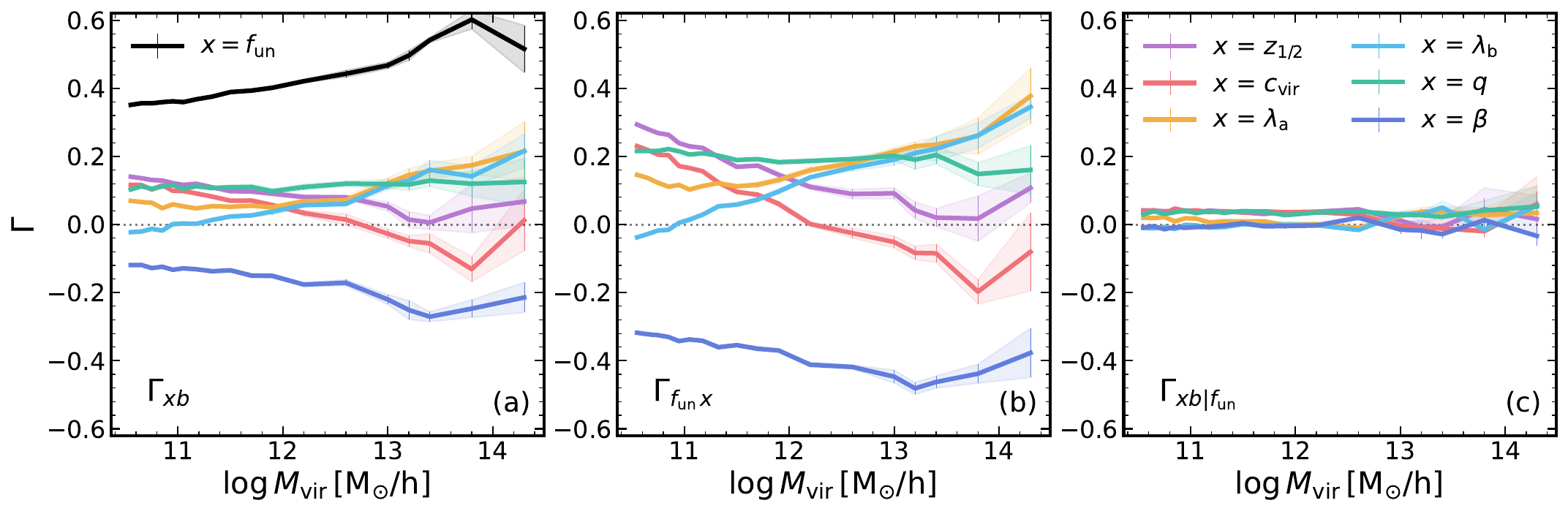}
    \caption{
    {\figem Spearman correlation coefficients between halo bias, $f_{\rm un}$ 
    and halo internal properties.}
    {\figem (a)}, correlations between halo bias $b$ and each internal property 
    $x$: formation time ($z_{1/2}$, purple), 
    concentration ($c_{\rm vir}$, red), 
    spins in two definitions ($\lambda_{\rm a}$, orange; 
    $\lambda_{\rm b}$, light-blue), 
    shape parameter ($q$, green) and velocity anisotropy ($\beta$, blue). 
    For comparison, the correlation between $b$ and $f_{\rm un}$ 
    is also shown (black).
    {\figem (b)} Correlations between $f_{\rm un}$ and the internal properties.
    {\figem (c)} Correlations between $b$ and the internal properties after controlling for $f_{\rm un}$.
    In each panel, shading and error bars show the bootstrap errors;
    the horizontal dotted line marks zero correlation.
    This figure shows that $f_{\rm un}$ is the strongest predictor of halo bias, 
    and that controlling for $f_{\rm un}$ largely removes the residual 
    biases of internal halo properties (see \S\ref{ssec:fun-bias} for details).
    }
    \label{fig:spearman_fun_delta}
\end{figure*}

\begin{figure*}
    \includegraphics[width=0.9\textwidth]{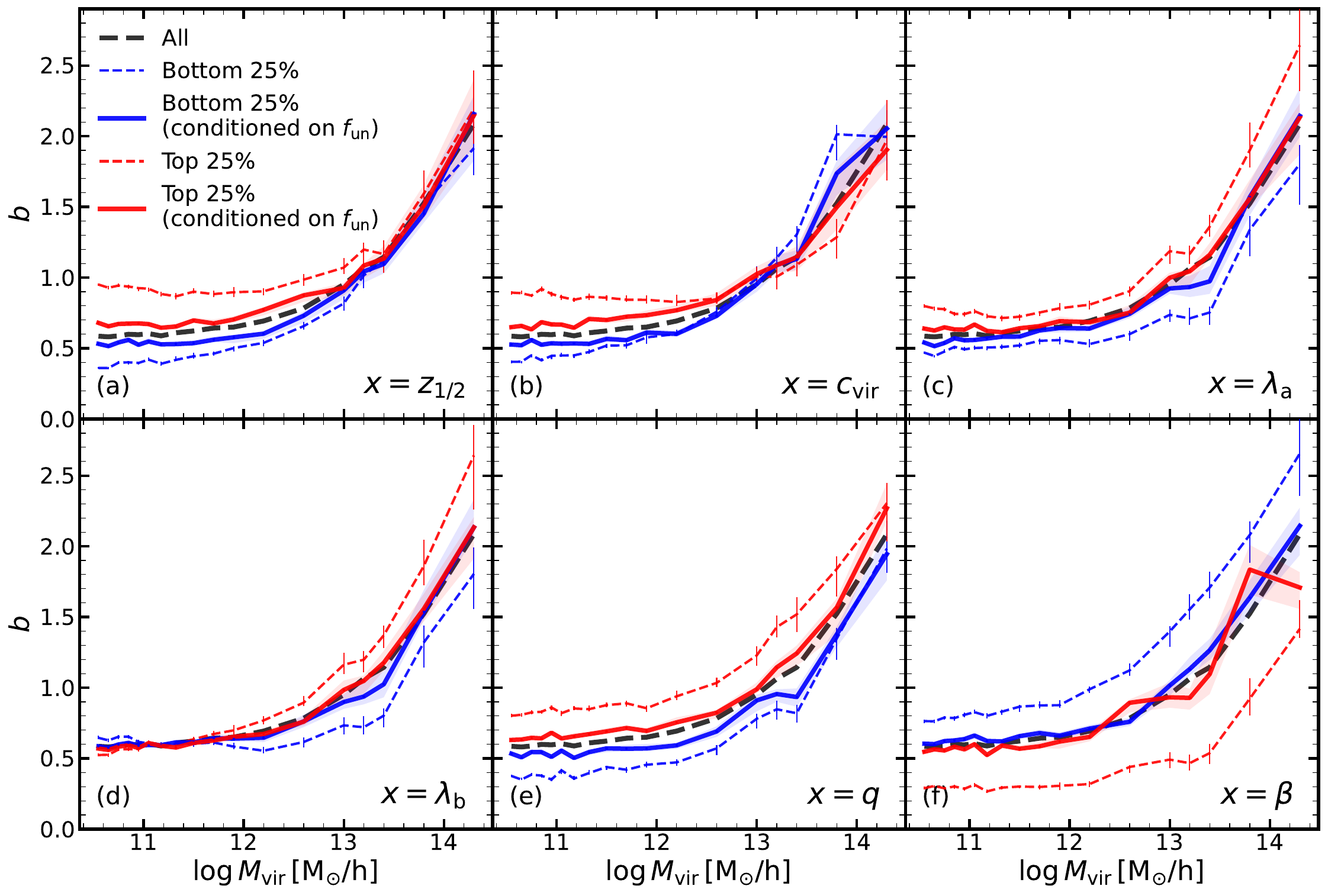}
    \caption{
    {\figem Halo secondary bias with or without controlling for $f_{\rm un}$.}
    {\figem (a)}--{\figem (f)}, average halo bias in bins of $M_{\rm vir}$ 
    for haloes in the top and bottom quartiles of each internal property, 
    $x$, as indicated in the bottom-right corner of each panel.
    In each panel, coloured dashed curves show the bias directly averaged
    from the haloes in the quartiles of $x$, while coloured solid curves show 
    the bias averaged with weights that align the $f_{\rm un}$ distributions 
    of each quartile to that of all haloes in the corresponding 
    $M_{\rm vir}$ bin. 
    Shaded regions or error bars associated with each curve show the bootstrap 
    errors.
    For reference, the black dashed curve shows the halo bias averaged over 
    all haloes.
    This figure confirms the result from correlation analyses (Fig.~\ref{fig:spearman_fun_delta}) that controlling for $f_{\rm un}$ largely 
    removes the secondary bias associated with internal halo properties
    (see \S\ref{ssec:fun-bias} for details). 
    %\todo[YC]{Check the colors
    %of the label.}
    }
    \label{fig:control_bias}
\end{figure*}

\section{Secondary bias of haloes}
\label{sec:secondary_bias}

\subsection{$f_{\rm un}$ as a mediator of secondary bias}
\label{ssec:fun-bias}

To see the relation between $f_{\rm un}$ and secondary bias of haloes, we partition
the simulated haloes into bins of $M_{\rm vir}$, and quantify the relation 
between $f_{\rm un}$ and halo bias within each mass bin. We divide the haloes in 
each $M_{\rm vir}$ bin into quartiles based on their $f_{\rm un}$ values,
and we compute the average bias, $b$ (Eq.~\ref{eq:bias}), among haloes for 
each quartile.
Fig.~\ref{fig:diffm_un_bias} shows average $b$ in each quartile as a function 
of $M_{\rm vir}$ (coloured curves), adopting four different radial ranges in 
the measurement of $f_{\rm un}$ (panels a--d). 
For reference, the average $b$ over all haloes, without binning $f_{\rm un}$, 
is shown by a black curve; a random subset of individual haloes is shown 
as scatter points, with their $f_{\rm un}$ colour-coded.

As expected, the primary bias, namely the relation between $b$ and $M_{\rm vir}$,
shows a positive slope with a flattening at low masses 
($M_{\rm vir}\lesssim 10^{12}\msun$). 
At a fixed $M_{\rm vir}$, haloes with a larger $f_{\rm un}$ are 
more strongly clustered, with low-mass haloes in the top quartile reaching a bias 
comparable to the mass-only bias of haloes with 
$M_{\rm vir} \approx 10^{14} \msun$.
Such a tendency for higher-$f_{\rm un}$ haloes to exhibit higher bias is
consistent with the difference between the environments in which the 
two example haloes reside (Fig.~\ref{fig:fun_example}b, c).
Notably, the $b$--$M_{\rm vir}$ relations for different $f_{\rm un}$ quartiles
are clearly separated over the entire range of $M_{\rm vir}$ examined here, and 
the separation remains pronounced in all radial ranges. 
The most pronounced $b$--$f_{\rm un}$ correlation
is observed for $f_{\rm un}$ measured within $2$--$3\,R_{\rm vir}$
(Fig.~\ref{fig:diffm_un_bias}c). In comparison, assembly bias measured 
with respect to other internal properties of haloes (e.g. those listed in 
\S\ref{ssec:halo_property}) is generally weaker and/or mass-dependent
(see e.g. \citealt{gaoAgeDependenceHalo2005}; see also the correlation 
analyses below).

The strong correlation between $f_{\rm un}$ and halo bias suggests an origin
of the secondary bias in which the environment shapes $f_{\rm un}$, and through it, 
the internal (structural and assembly) properties of haloes.
If $f_{\rm un}$ indeed serves as such a mediator between environment and haloes, 
this immediately yields a testable prediction: the correlation between 
bias and any internal property of haloes should vanish, once $f_{\rm un}$ is 
fixed. 
% \gamma looks too small in the rendered PDF compared with the subscripts. 
% We switch to \Gamma, which looks bigger. 
% Use z for latent or mediating variable is also common.
A common approach to quantify such a dependence graph is through 
the conditional correlation coefficient, $\Gamma_{x y \mid z}$,
defined as \citep[e.g.][]{ramakrishnanCosmicWebAnisotropy2019}:
\begin{equation}
\Gamma_{xy \mid z} = \frac{\Gamma_{xy}-\Gamma_{zy} \Gamma_{zx}}
{\sqrt{(1-\Gamma_{zy}^2) (1-\Gamma_{zx}^2)}},
\label{eq:conditional_correlation}
\end{equation}
for variables $x$ and $y$ conditioned on the mediating variable
$z$, where $\Gamma_{xy}$, $\Gamma_{zy}$ and $\Gamma_{zx}$ are the 
(unconditional) pairwise Spearman correlation coefficients among
the variables. A value of $\Gamma_{xy \mid z}$ close to zero 
indicates that the correlation between $x$ and $y$ is largely accounted for 
by their correlation with $z$, implying conditional independence 
between $x$ and $y$ given $z$.
In our application, $z$ is the unbound fraction ($f_{\rm un}$), $y$ is 
the bias factor ($b$), and $x$ is one of the internal properties of the halo
(e.g. $z_{1/2}$, $c_{\rm vir}$, $\lambda_{\rm a}$, $\lambda_{\rm b}$,
$q$ or $\beta$; see \S\ref{ssec:halo_property} for definitions). 
Unless explicitly stated, hereafter, we use $f_{\rm un}$ in the radial 
range of $2$--$3\,R_{\rm vir}$ as our default choice, 
where the $b$--$f_{\rm un}$ correlation is strongest
(Fig.~\ref{fig:diffm_un_bias}c; see also Appendix~\ref{app:radial-range-of-unbound-particles} for alternative 
apertures). 
The uncertainty for the conditional correlation coefficient is estimated as 
the standard deviation among 50 bootstrap resamples with replacement.

Fig.~\ref{fig:spearman_fun_delta}(a) shows $\Gamma_{xb}$, the correlation between 
each internal property $x$ and halo bias $b$, which quantifies the secondary 
bias that has been discovered in previous studies.
Older haloes (with higher $z_{1/2}$) are clustered more strongly (exhibiting 
higher $b$) at the low-mass end, but this trend disappears at the high-mass end
\citep[e.g.,][]{gaoAgeDependenceHalo2005}.
$c_{\rm vir}$ behaves similarly to $z_{1/2}$ at the low-mass end, where
the halo profile is found to be closely tied to the formation history
\citep{wechslerConcentrationsDarkHalos2002,Ludlow2013MassProfileAccretion}, 
but the sign of the correlation reverses at the high-mass end
\citep[e.g.,][]{gaoAssemblyBiasClustering2007}, a consequence likely
arising from the configuration of the protohalo determined by the 
distribution of progenitor haloes beyond that in the main branch
\citep{maoAssemblyBiasExploring2018,wangConcentrationsDarkHaloes2020,wangCharacterizingAssemblyDark2024}, leading to a delayed response for the formation time (see discussion in Appendix~\ref{app:internal-properties-in-alpha-qe-plane}). 
Haloes with lower $\lambda_{\rm b}$ are more strongly clustered at the 
low-mass end, a trend that reverses at $M_{\rm vir} \gtrsim 10^{11.5}\msun$
\citep[e.g.,][]{sato-politoDependenceHaloBias2019}.
This inversion of $\Gamma_{\lambda_b b}$ is sensitive to the set of 
particles in use: additionally including unbound particles produces 
a correlation $\Gamma_{\lambda_a b}$ with a positive sign over the entire
mass range \citep{MaInterpretingStrongClustering2026}, discussed further in Appendix~\ref{app:internal-properties-in-alpha-qe-plane}.
Rounder haloes, and haloes with more tangentially biased orbits, 
are clustered more strongly over the entire mass range 
\citep[e.g.,][]{Faltenbacher2010AssemblyBias}.
For comparison, we show $\Gamma_{f_{\rm un}b}$, the correlation between 
the unbound fraction $f_{\rm un}$ and halo bias $b$, which maintains 
a positive sign and a stable magnitude of $0.35$--$0.6$
across the entire mass range examined here.
Importantly, $\Gamma_{f_{\rm un}b}$ is much larger than all the other 
$\Gamma_{x b}$, reinforcing the significance of the unbound fraction 
in tracing secondary bias.

The significance of $\Gamma_{f_{\rm un}b}$, together with the constancy of 
its sign and magnitude across halo mass, yields a clear expectation: if
$f_{\rm un}$ indeed mediates the effect of the environment on an internal
property of a halo, $x$, then 
\begin{enumerate}
    \item $f_{\rm un}$ and $x$ should exhibit a significant correlation, 
    for any $x$ that exhibits significant secondary bias;
    \item the sign of $\Gamma_{f_{\rm un} x}$, and the mass dependence of
    its sign and magnitude, should be similar to those of $\Gamma_{x b}$;
    \item the conditional correlation, $\Gamma_{x b \mid f_{\rm un}}$, 
    should vanish.
\end{enumerate}

Fig.~\ref{fig:spearman_fun_delta}(b) and (c) show $\Gamma_{f_{\rm un} x}$ and
$\Gamma_{x b \mid f_{\rm un}}$ for various internal properties of haloes, 
confirming the expectations listed above:
all the internal properties exhibit a significant correlation with 
$f_{\rm un}$; the sign of $\Gamma_{f_{\rm un} x}$ is the same as that of 
$\Gamma_{x b}$, and for $x = c_{\rm vir}$ and $\lambda_{\rm b}$,
the sign of $\Gamma_{f_{\rm un} x}$ is inverted at nearly the same halo mass 
as for $\Gamma_{x b}$; the magnitude of $\Gamma_{f_{\rm un} x}$ also 
closely follows that of $\Gamma_{x b}$, albeit higher by a factor of 
$\approx 2$--$3$; all the conditional correlations, 
$\Gamma_{x b \mid f_{\rm un}}$, are close to zero 
(with magnitude $\lesssim 4\%$) over the entire mass range examined 
here. Notably, the correlation between velocity anisotropy ($\beta$)
and $b$ almost completely disappears after conditioning on $f_{\rm un}$,
suggesting that $f_{\rm un}$ outperforms other environmental
indicators, such as the tidal anisotropy $\alpha$
\citep{ramakrishnanCosmicWebAnisotropy2019}, in accounting for the velocity
structure of haloes (see \S\ref{ssec:tidal-anisotropy}
and Appendix~\ref{app:alpha-bias} for more discussion).
The most obvious residual, at the level of $\Gamma_{x b \mid f_{\rm un}} \approx 0.04$,
is seen only for $z_{1/2}$, $c_{\rm vir}$ and $q$.
These three internal properties
are more susceptible to peculiar events, such as `splashbacks' \citep[e.g.][]{liAssemblyBiasDwarfsizedDark2013},
that take place long ago in the histories of the haloes, 
and are thus less completely tracked by the mediator, $f_{\rm un}$, 
which is measured at $z = 0$ (see \S\ref{ssec:splashback} for quantitative discussion).
% whereas spin and velocity anisotropy respond to recent accretion and mergers; 
% $f_{\rm un}$ thus captures only part of that history, leaving minor residuals 
% for the long-term properties.

To verify that these conclusions are not an artifact of the 
limited ability of correlation analysis to capture non-linear relations, we 
also adopt a conventional metric of secondary bias.
We take the top and bottom quartiles of the distribution of each internal 
property in each bin of $M_{\rm vir}$, compute the average $b$ for haloes in 
each quartile, and use the difference of the average $b$ between the two 
quartiles as the metric of (unconditional) secondary bias. To obtain the 
secondary bias conditioned on $f_{\rm un}$ 
we assign a weight to each halo 
in a quartile so that the weighted distribution of $f_{\rm un}$ for haloes in 
that quartile is identical to that of all haloes within the same bin of $M_{\rm vir}$ 
\citep[see e.g.][for a similar weighting scheme]{zhangUnexpectedClusteringPattern2025}; 
the average of $b$ using these weights is then computed for haloes in each 
quartile, and the difference between the two quartiles measures
the secondary bias conditioned on $f_{\rm un}$.

Fig.~\ref{fig:control_bias} shows the unconditional (dashed curves) and 
conditional (solid curves) secondary biases.
%\todo{with bootstrap error bars (50 resamples)??? where are the errorbars}. 
As expected, all six internal properties show significant
unconditional secondary bias; the conditional ones, however, 
nearly vanish for $\lambda_{\rm a}$, $\lambda_{\rm b}$, 
and $\beta$, and are largely reduced for the other properties as well. 
These results support the role of $f_{\rm un}$ as a mediator
in propagating the environmental influence on internal properties of haloes.

Physically, the correlation between $b$ and internal properties of haloes also 
suggests a causal direction, since the large-scale environment is not likely to 
respond directly to the evolution of a halo. Thus, our correlation analysis, 
together with the physically implied causal direction, 
points to a picture in which a denser environment induces a higher unbound 
fraction around haloes, and such an increase in the unbound fraction 
subsequently influences the assembly and structure of haloes.
In what follows, we compare our findings with those obtained using 
alternative environmental metrics. This places our findings in a broader 
context, and helps clarify how the large-scale environment affects
haloes through $f_{\rm un}$.
% We emphasize that these diagnostics concern conditional dependence rather than the direction of 
% causation: a vanishing conditional correlation establishes that $f_{\rm un}$ organizes the 
% secondary-bias signal, while the physical picture behind this organization is examined in .

\begin{figure}
    \includegraphics[width=0.48\textwidth]{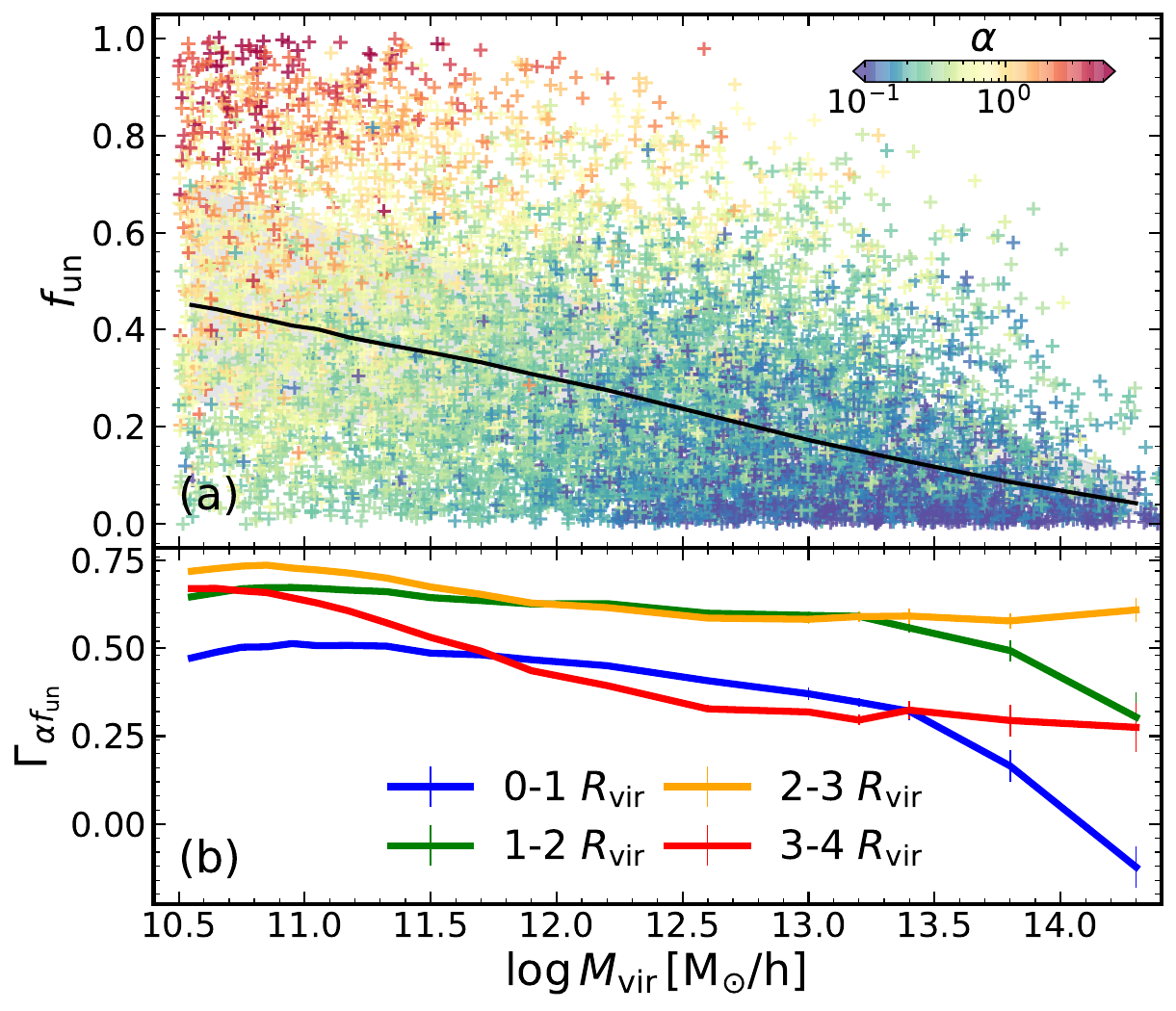}
    \caption{
    {\figem The relation between unbound mass fraction $f_{\rm un}$ 
    and tidal anisotropy ($\alpha$).}
    {\figem (a)},  $f_{\rm un}$ (measured within $2$--$3\,R_{\rm vir}$) versus 
    halo mass for 10,000 randomly selected individual haloes (plus markers),
    coloured by $\alpha$. 
    The solid curve shows the median $f_{\rm un}$ of haloes
    in bins of $M_{\rm vir}$, and the shaded region
    shows the $25^{\rm th}$--$75^{\rm th}$ percentiles.
    {\figem (b)}, Spearman correlation coefficients between $f_{\rm un}$ and 
    $\alpha$ in bins of halo mass, with $f_{\rm un}$ measured within different 
    radial ranges (coloured curves).
    This figure shows that $f_{\rm un}$ is strongly correlated with 
    $\alpha$, indicating a role of $f_{\rm un}$ as a mediator of the 
    environmental effects on haloes
    (see \S\ref{ssec:tidal-anisotropy} for details).
    }
    \label{fig:fun_alpha}
\end{figure}

\begin{figure*}
    \includegraphics[width=0.9\textwidth]{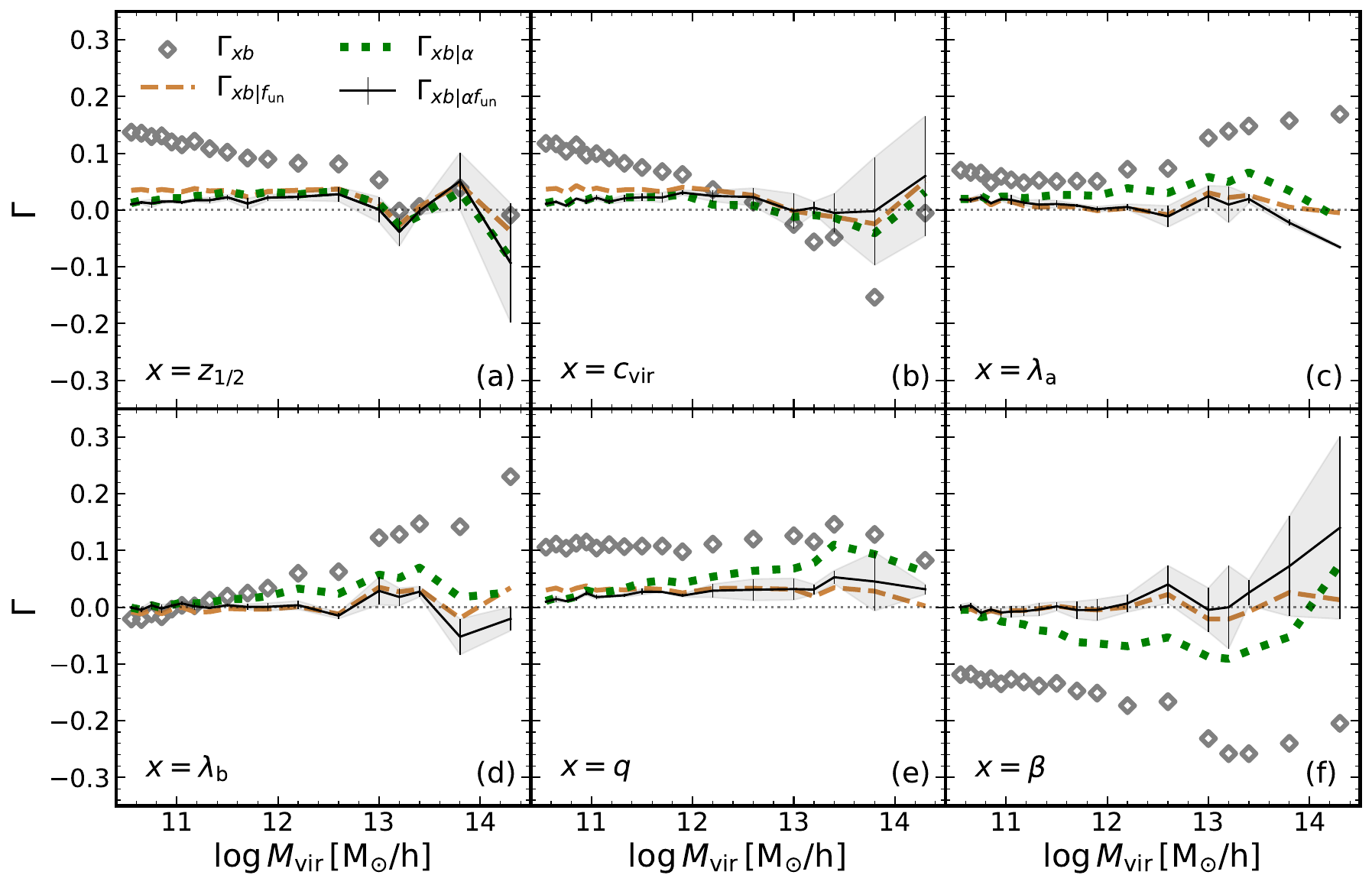}
    \caption{
    {\figem Spearman correlation coefficients between bias $b$ and 
    internal properties of haloes with different conditioning 
    variables.} 
    Each panel shows the correlation between $b$ and one of the six internal
    properties of haloes.
    Results using four combinations of conditioning variables are shown: 
    no conditioning (grey diamonds), 
    conditioned on $f_{\rm un}$ (orange dashed),
    $\alpha$ (green dotted), or both (black solid).
    For clarity, we show only the bootstrap errors using shading and error bars 
    for the joint conditioning.
    The horizontal grey dotted line indicates zero correlation.
    This figure shows that conditioning on either $f_{\rm un}$ or 
    $\alpha$ removes most of the secondary bias, while conditioning on both 
    leads to little improvement, supporting the role of 
    $f_{\rm un}$ as a mediator of the effects of environment
    characterized by $\alpha$ onto haloes
    (see \S\ref{ssec:tidal-anisotropy} for details).
    }
    \label{fig:spearman_alpha_fun}
\end{figure*}

\subsection{Linking $f_{\rm un}$ to alternative environmental metrics}
\label{ssec:related-to-previous-mechanism}

Previous studies have proposed explanations for secondary bias by linking 
the internal properties of haloes to alternative indicators of
environment. These studies identify a key factor, the tidal field,
that can alter the assembly and structure of haloes and thereby generate the
secondary bias. Descriptors for the tidal field include direct measurements of the tidal
tensor \citep[e.g.][]{hahnPropertiesDarkMatter2007,ForeroRomero2014CosmicWebAlignments}, 
metrics derived from the tidal tensor, such as its eigenvalues \citep{wangInternalPropertiesEnvironments2011} and anisotropy ($\alpha$, see 
\S\ref{ssec:halo_property}), and indirect indicators such as whether
or not a halo has experienced a splashback event in the past \citep[e.g.][]{wangDistributionEjectedSubhalos2009,liAssemblyBiasDwarfsizedDark2013,sunayamaScaledependenceHaloAssembly2016,mansfieldThreeCausesLowmass2020,tucciPhysicalOriginsLowmass2021}.
As we will show in the following, these indicators are not only linked
to $f_{\rm un}$, providing complementary viewpoints on how environment affects 
haloes, but also account for the residual of secondary bias that is missed by 
$f_{\rm un}$, particularly that involving the three internal properties 
($z_{1/2}$, $c_{\rm vir}$, $q$) whose residual is small 
but remains noticeable after controlling for $f_{\rm un}$ 
(at a level of $\Gamma_{xb \mid f_{\rm un}} \approx 4\%$; see Fig.~\ref{fig:spearman_fun_delta}c).

\subsubsection{Tidal anisotropy}
\label{ssec:tidal-anisotropy}

As shown by \citet{paranjapeHaloAssemblyBias2018} and \citet{Paranjape2021AnalyticalHaloModels}, the tidal anisotropy ($\alpha$), which quantifies the asymmetry in the gravitational environment
of a halo, can effectively distinguish between different cosmic-web environments
the halo resides in: an isotropic environment, such as voids and nodes, yields a small
$\alpha$, while an anisotropic environment, in particular that around
massive filaments and sheets, yields a large $\alpha$;
a value of $\alpha \lesssim 0.5$ and $\gtrsim 1$ can indicate
highly isotropic and anisotropic regions, respectively 
\citep[e.g.][]{paranjapeHaloAssemblyBias2018}. 
Most relevant to this study, $\alpha$ has been proposed as the driving factor that 
determines the assembly and structure of haloes at a given mass, 
while the secondary bias, namely the correlation between the large-scale 
bias factor ($b$) and internal properties of haloes, only emerges due to the 
correlation between $\alpha$ and $b$ \citep{paranjapeHaloAssemblyBias2018,ramakrishnanCosmicWebAnisotropy2019}.
This is very similar to the role of $f_{\rm un}$ in controlling the secondary 
bias (\S\ref{ssec:fun-bias}), suggesting that higher $f_{\rm un}$ tends 
to be seen in more anisotropic environments, and vice versa.

In Fig.~\ref{fig:fun_alpha}(a), we show the distribution of individual haloes in the 
$f_{\rm un}$--$M_{\rm vir}$ plane, with colour indicating the tidal anisotropy
$\alpha$.
Massive haloes ($M_{\rm vir} > 10^{13} \msun$) exhibit low $\alpha$ values, 
consistent with their preference to reside within node environments \citep[e.g.][]{paranjapeHaloAssemblyBias2018}. 
In contrast, lower-mass haloes exhibit systematically higher $\alpha$ values,
as well as much wider scatter at fixed mass, reflecting the 
diverse environments they inhabit, as already evidenced by the two examples 
shown in Fig.~\ref{fig:fun_example}(b) and (c). Haloes with higher 
$f_{\rm un}$ tend to be found in more anisotropic environments (showing higher 
$\alpha$ values), and such a correlation is observed across the 
range of $M_{\rm vir}$ examined here.
Lower-mass haloes appear to exhibit stronger correlations between 
$f_{\rm un}$ and $\alpha$, 
and in particular, haloes with $f_{\rm un} \gtrsim 50\%$ in their neighbouring 
regions of $2$--$3\,R_{\rm vir}$ often show $\alpha \gtrsim 1$.
Fig.~\ref{fig:fun_alpha}(b) provides a quantitative confirmation:
a positive correlation coefficient $\Gamma_{\alpha f_{\rm un}}$ is present 
across the entire mass range examined.
Adopting $f_{\rm un}$ measured within $2$--$3\,R_{\rm vir}$,
$\Gamma_{\alpha f_{\rm un}}$ is consistently high across the entire mass range,
with haloes at the low-mass (high-mass) end showing slightly stronger (weaker) 
correlations of $\Gamma_{\alpha f_{\rm un}} \approx 0.75$ ($0.6$).
While the overall trends remain consistent across different 
radial ranges, $f_{\rm un}$ measured within $2$--$3\,R_{\rm vir}$ shows the 
strongest correlation with $\alpha$ in nearly all mass bins,
providing further justification for using $f_{\rm un}$ measured within 
$2$--$3\,R_{\rm vir}$ as our default indicator of halo 
environment.

The consistently high $f_{\rm un}$--$\alpha$ correlation (adopting $f_{\rm un}$ 
measured within $2$--$3\,R_{\rm vir}$), together with the role of $f_{\rm un}$ 
in controlling the secondary bias, immediately suggests two implications: 
(i) the pattern of the correlation between $\alpha$
and an internal property $x$ of a halo mirrors that between $f_{\rm un}$
and this internal property (i.e. $\Gamma_{\alpha x}$ and $\Gamma_{f_{\rm un} x}$ 
should share the same sign and mass dependence);
(ii) conditioning on either $f_{\rm un}$ or $\alpha$ should largely account 
for the secondary bias, while conditioning on both should provide little 
incremental suppression of the secondary bias.
In Appendix~\ref{app:alpha-bias} and Fig.~\ref{fig:spearman_alpha_delta}, 
we examine the correlation between $\alpha$ and halo secondary properties 
and find results consistent with expectation (i).
To examine the expectation (ii), we consider the chain of correlation 
coefficients with incremental conditioning: 
$\Gamma_{b x}$, the unconditional correlation; 
$\Gamma_{b x \mid f_{\rm un}}$ and $\Gamma_{b x \mid \alpha}$, the 
correlation conditioned on $f_{\rm un}$ and $\alpha$, 
respectively; $\Gamma_{b x \mid \alpha f_{\rm un}}$, the correlation
conditioned on both $f_{\rm un}$ and $\alpha$, evaluated as
\citep{kendallRankCorrelationMethods1990}:
\begin{equation}
\Gamma_{bx \mid \alpha f_{\rm un}} =
    \frac{
        \Gamma_{bx \mid f_{\rm un}} 
        - \Gamma_{\alpha b \mid f_{\rm un}}\Gamma_{\alpha x \mid f_{\rm un}}}
    {\sqrt{
        \left(1-\Gamma_{\alpha b \mid f_{\rm un}}^2\right)
        \left(1-\Gamma_{\alpha x \mid f_{\rm un}}^2\right)
    }}
\label{eq:conditional_alpha_Q}
\end{equation}

In Fig.~\ref{fig:spearman_alpha_fun}, we show the correlation coefficients with 
incremental conditioning in bins of $M_{\rm vir}$.
Compared with the unconditional correlations, conditioning on either $f_{\rm un}$ 
or $\alpha$ substantially reduces secondary bias.
The performance of conditioning depends on the specific halo property 
under consideration: $f_{\rm un}$ alone almost totally eliminates
the secondary bias in spins ($\lambda_{\rm a}$ and $\lambda_{\rm b}$) 
and velocity anisotropy ($\beta$), as already demonstrated 
in Figs.~\ref{fig:spearman_fun_delta} and 
\ref{fig:control_bias}; $\alpha$, however, outperforms $f_{\rm un}$ 
in controlling the secondary bias in formation time ($z_{1/2}$), 
concentration ($c_{\rm vir}$), and axis ratio ($q$; 
at $M_{\rm vir} \lesssim 10^{11}\msun$); conditioning on 
both provides additional reduction in the residual
of secondary bias, but the improvement over conditioning on
$f_{\rm un}$ alone is modest.
We caution, however, that the exact value of a correlation coefficient
involving $\alpha$ depends on the definitions adopted for the properties 
being examined (see Appendix~\ref{app:alpha-bias} for details).

The consistently high correlation between $f_{\rm un}$ and $\alpha$, and their 
consistent performance in controlling secondary bias, suggest a causal
direction, $\alpha \rightarrow f_{\rm un} \rightarrow x$: 
a tidal field with larger anisotropy leads to stronger shear motion 
between the halo and its circum-halo medium, leaving a smaller fraction of the mass to be bound to, and accreted by the halo.
In this scenario, tidal anisotropy is the cause, unbound matter is the 
mediator, and the variation in the assembly and structure of haloes is 
the outcome. In \S\ref{sec:physics-of-fun-vs-b}, we will examine the underlying 
processes traced by $f_{\rm un}$ and show 
that such a causal relationship holds, and only holds, for a subset of 
haloes that `co-move' with their circum-halo medium.

%Conditioning jointly on $\alpha$ and an outer-shape measure of the halo ($Q$) leaves residuals for the dynamical properties nearly as small as those obtained with $f_{\rm un}$ alone (grey dashed error bars in Fig.~\ref{fig:spearman_alpha_fun}) (see more detail in \S\ref{ssec:merger-response}).
%We note that our results for $\alpha$ differ slightly from those of \cite{ramakrishnanCosmicWebAnisotropy2019}. This minor discrepancy may stem from differing definitions of the secondary properties, bias, and tidal anisotropy, a point we discuss in more detail in Appendix~\ref{app:alpha-bias}. 

\begin{figure}
    \includegraphics[width=0.45\textwidth]{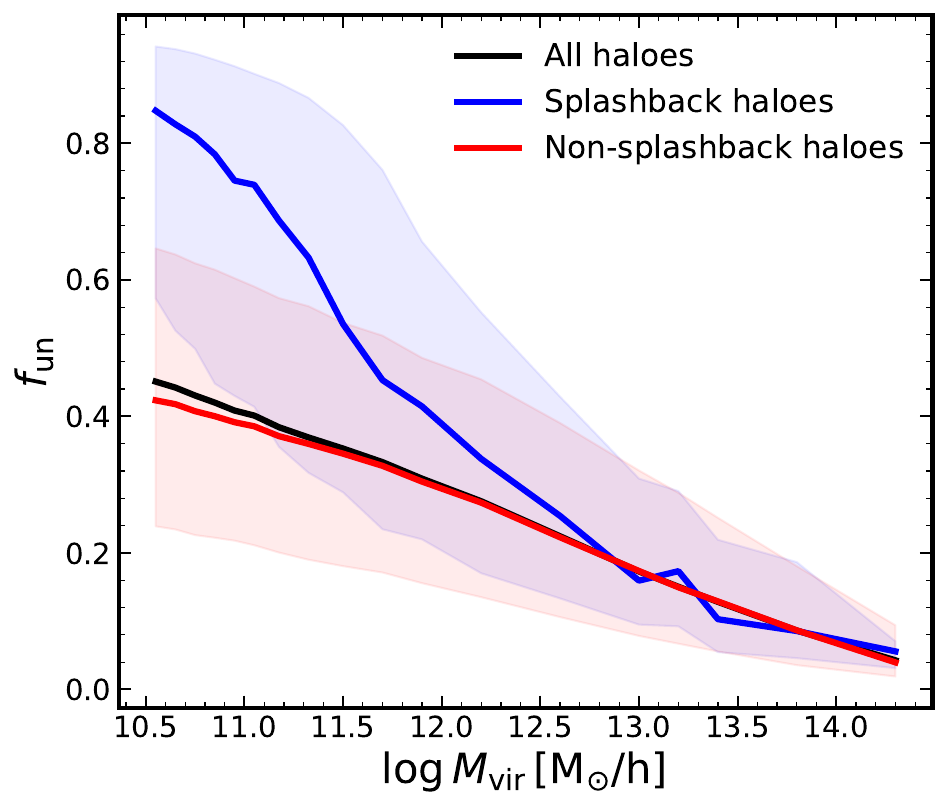}
    \caption{
    {\figem $f_{\rm un}$ versus halo mass for splashback and 
    non-splashback haloes.} 
    Solid curves show the medians of $f_{\rm un}$, 
    with shaded regions indicating the $25^{\rm th}$--$75^{\rm th}$ percentiles,
    for splashback (blue), non-splashback (red) and all haloes (black)
    in bins of halo mass. This figure shows that splashback haloes have 
    systematically higher $f_{\rm un}$ (see \S\ref{ssec:splashback} for details).
    }
    \label{fig:splashback_fun}
\end{figure}

\begin{figure*}
    \begin{minipage}
    {0.9\textwidth}
        \centering
        \includegraphics[width=0.95\textwidth]{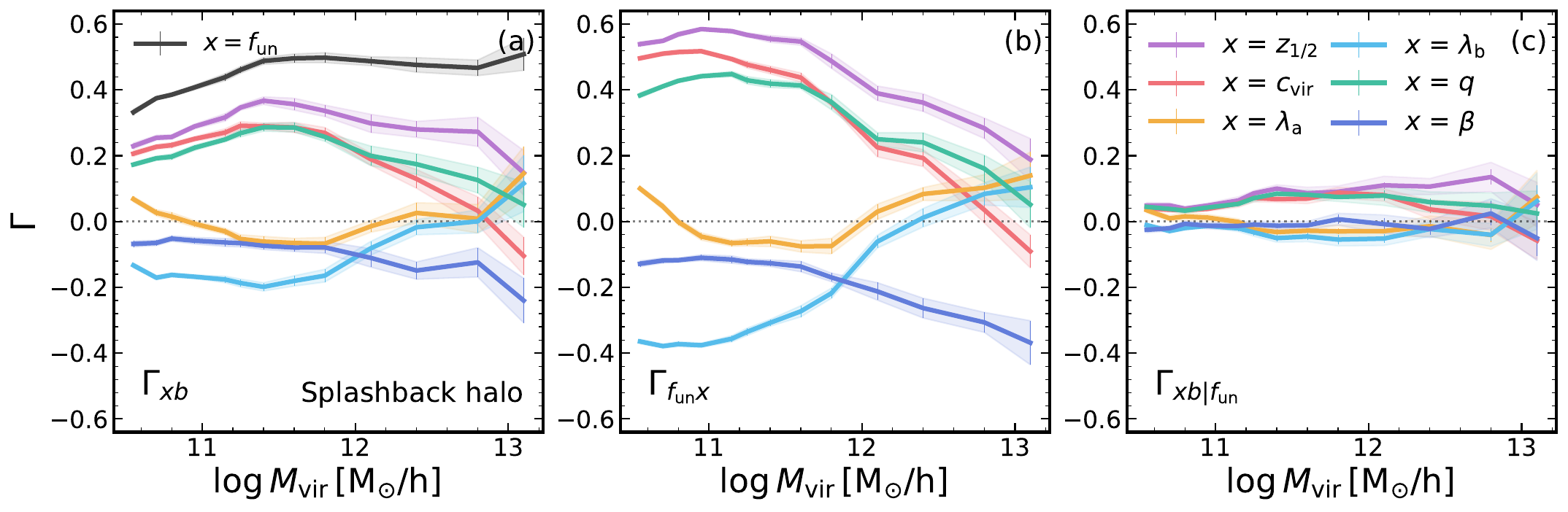}
    \end{minipage}
    \begin{minipage}
        {0.9\textwidth}
            \centering
            \includegraphics[width=0.95\textwidth]{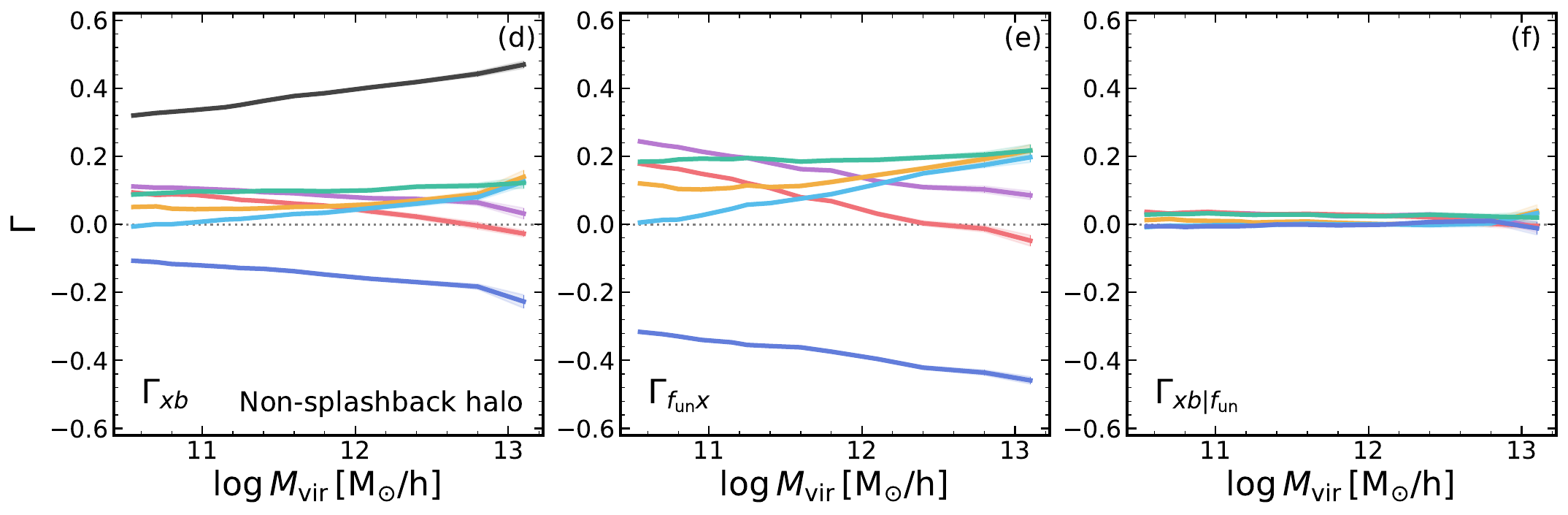}
    \end{minipage}
    \caption{
        {\figem Comparison of the Spearman correlation coefficients for 
        splashback and non-splashback haloes.}
        {\figem (a)}--{\figem (c)}, unconditional and conditional Spearman 
        correlation coefficients similar to Fig.~\ref{fig:spearman_fun_delta}, 
        but for splashback haloes only.
        {\figem (d)}--{\figem (f)}, similar but for non-splashback haloes only.
        In each panel, the horizontal dotted line indicates zero correlation.
        This figure shows that controlling $f_{\rm un}$ reduces 
        secondary bias for both splashback and non-splashback haloes, 
        although the residual bias is more pronounced for the splashback
        haloes (see \S\ref{ssec:splashback} for details).
    }
    \label{fig:spearman_fun_delta_splashback}
\end{figure*}

\subsubsection{Splashback events}
\label{ssec:splashback}

A splashback event occurs when a halo, after an excursion as a satellite subhalo
of another halo, is later ejected \citep[e.g.][]{Gill2005EvolutionSubstructure,Ludlow2009TheUnorthodoxOrbits}. The population of splashback haloes 
is found to be partly responsible for the secondary bias, particularly 
for those with low $M_{\rm vir}$
\citep[e.g.][]{wangDistributionEjectedSubhalos2009,liAssemblyBiasDwarfsizedDark2013,sunayamaScaledependenceHaloAssembly2016,mansfieldThreeCausesLowmass2020,tucciPhysicalOriginsLowmass2021}. 
As with $\alpha$, it is therefore expected that $f_{\rm un}$ also 
correlates with the occurrence of splashback events in the 
histories of haloes, and such a correlation may partly explain the 
secondary bias introduced by the splashback population.
To quantify this, we identify all splashback events for each halo in our 
sample by tracing back in time along the main progenitor branch of the merger
tree rooted in the central subhalo of the halo in question, and we define 
the splashback redshift, $z_{\rm sb}$, of the halo 
as the last redshift at which the halo was a satellite subhalo. 
Since a splashback event at too early a redshift may not be relevant for the 
current state of the halo, we limit the range of $z_{\rm sb}$
by defining haloes with $z_{\rm sb} < 1$ as splashback
haloes; the remaining haloes are classified as non-splashback.
This criterion classifies $\approx 10\%$ of haloes at the low-mass 
end as splashback haloes.
% Because \texttt{SUBLINK} assigns subhalo status whenever a halo lies within the FOF group of its host, a looser criterion than the SO definition based on the virial radius, so this flag is easier to meet; 

Fig.~\ref{fig:splashback_fun} shows $f_{\rm un}$ as a function of 
$M_{\rm vir}$, separately for the splashback and non-splashback 
populations. At $M_{\rm vir} \gtrsim 10^{13}\msun$, 
the difference between the two populations in $f_{\rm un}$ is negligible, 
while at $M_{\rm vir} \lesssim 10^{10.5}\msun$, the median $f_{\rm un}$ of 
splashback haloes is about twice that of non-splashback haloes.
Such a difference may be expected: a lower-mass halo is more likely to be 
subdominant in gravity among the structures in its local environment, 
making its loosely bound exterior more susceptible to being stripped 
away once it makes an excursion around or into another structure.

To measure the correlations with the effects of splashback controlled,
we repeat the correlation analyses for Fig.~\ref{fig:spearman_fun_delta}
separately for the splashback and non-splashback populations,
and we show the resulting correlation coefficients in Fig.~\ref{fig:spearman_fun_delta_splashback}.
Consistent with the result of \citet{tucciPhysicalOriginsLowmass2021},
excluding splashback haloes removes the $\lambda_{\rm b}$-$b$ correlation
at the low-mass end ($M_{\rm vir} \approx 10^{10.5} \msun$).
The correlation coefficients $\Gamma_{xb}$ of the splashback population 
are larger in amplitude than those of the non-splashback population for 
properties $x = z_{1/2}$, $c_{\rm vir}$, and $q$ across the entire
mass range, and for $x = \lambda_{\rm b}$ at $M_{\rm vir} \lesssim 10^{12}\msun$, but not for $\lambda_{\rm a}$ and $\beta$ (Fig.~\ref{fig:spearman_fun_delta_splashback}a, d), indicating that splashback accounts 
for a portion, but not all, of the secondary bias.
The correlation coefficient $\Gamma_{f_{\rm un} x}$ (Fig.~\ref{fig:spearman_fun_delta_splashback}b, e) gives some clues about why 
the effect of splashback on secondary bias depends on the halo property 
in question: the correlation coefficients $\Gamma_{f_{\rm un} x}$ 
for $z_{1/2}$, $c_{\rm vir}$, and $q$ are much higher 
for the splashback population than for the non-splashback population; 
in contrast, $\Gamma_{f_{\rm un} \beta}$ is not enhanced, 
and in fact shrinks for the splashback population.

As shown in Fig.~\ref{fig:spearman_fun_delta_splashback}(c) and (f), 
the correlation coefficients after controlling for $f_{\rm un}$ are largely 
reduced for both the splashback and non-splashback populations.
As with the tidal anisotropy ($\alpha$), this suggests
a causal direction, $\text{splashback} \rightarrow f_{\rm un} \rightarrow x$: 
splashback events alter the fractions of unbound matter surrounding 
the haloes, and this matter in turn affects the assembly and structure
of these haloes. The mediator role of $f_{\rm un}$ is again implied in this 
causal chain, which, together with the chain $\alpha \rightarrow f_{\rm un} \rightarrow x$ inferred in \S\ref{ssec:tidal-anisotropy},
suggests that $f_{\rm un}$ is a sensitive tracer of physical processes 
operating at different spatial and temporal scales around haloes.
In the following, we examine these processes in more detail.

\begin{figure}
    \includegraphics[width=\columnwidth]{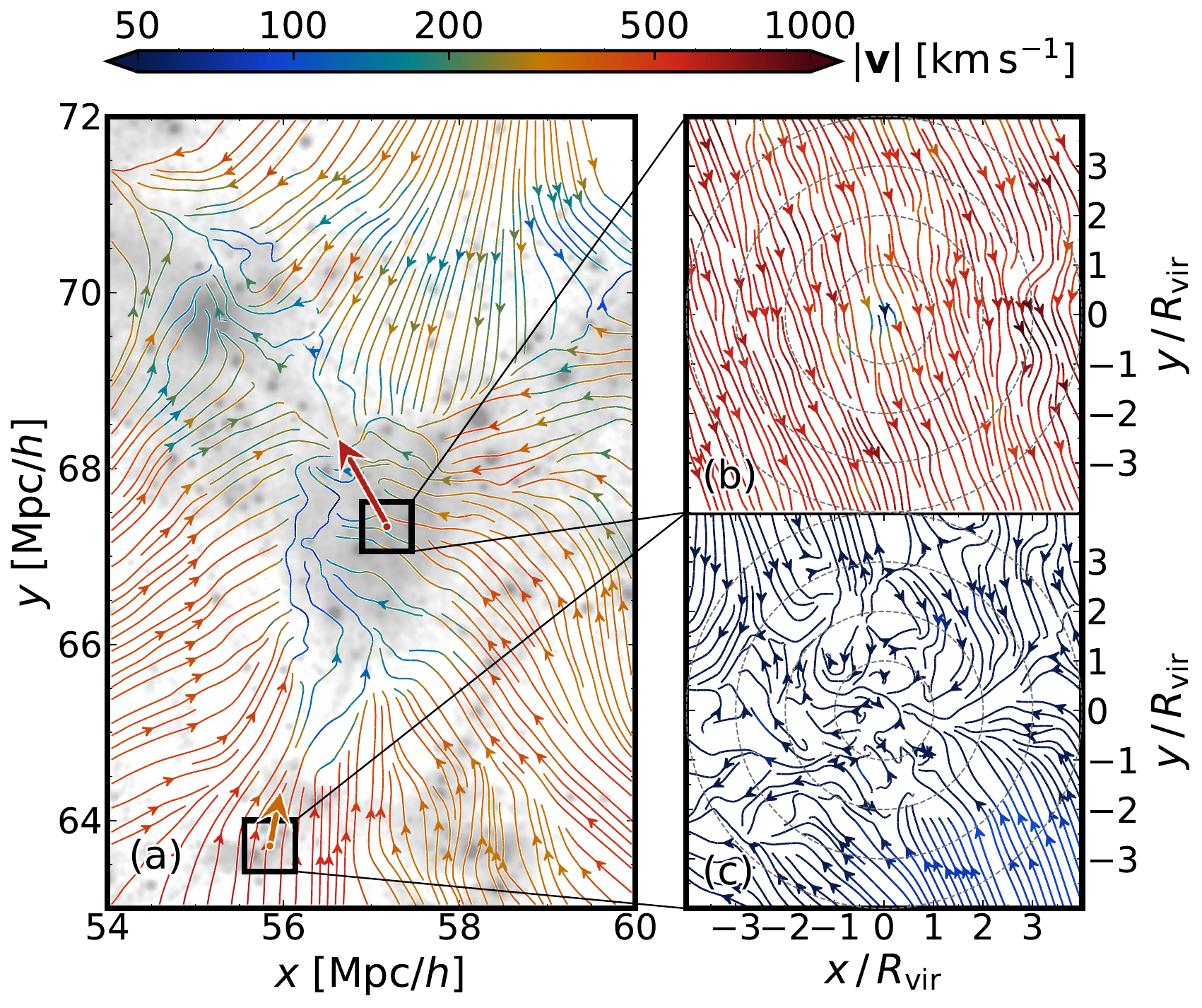}
    \caption{
    {\figem Velocity fields around the two example haloes.}
    Here we show the projected velocity field using streamlines
    in the same slice (panel {\figem a}), and around the same two example haloes 
    with extreme $f_{\rm un}$ values
    (panels {\figem b, c}), as those in Fig.~\ref{fig:fun_example}.
    The surface density field (grey background in panel a) is the same as that 
    in Fig.~\ref{fig:fun_example}(a).
    Velocity in panel {\figem (a)} is the peculiar velocity obtained from 
    particles in the simulation frame, and in panels {\figem (b)} and 
    {\figem (c)} is physical velocity (peculiar plus Hubble flow) 
    relative to the example halo. The peculiar velocities of the example haloes 
    are marked by arrows in panel {\figem (a)}. 
    Both streamlines and arrows are colour-coded by the magnitude of the
    projected velocity, according to the colour bar on the top.
    This figure reveals the distinct flow patterns around haloes with 
    different $f_{\rm un}$ values (see \S\ref{ssec:fun-tracing-flow} 
    for details).
    }
    \label{fig:large_scale_velocity_stream}
\end{figure}

\begin{figure*}
    \includegraphics[width=0.98\textwidth]{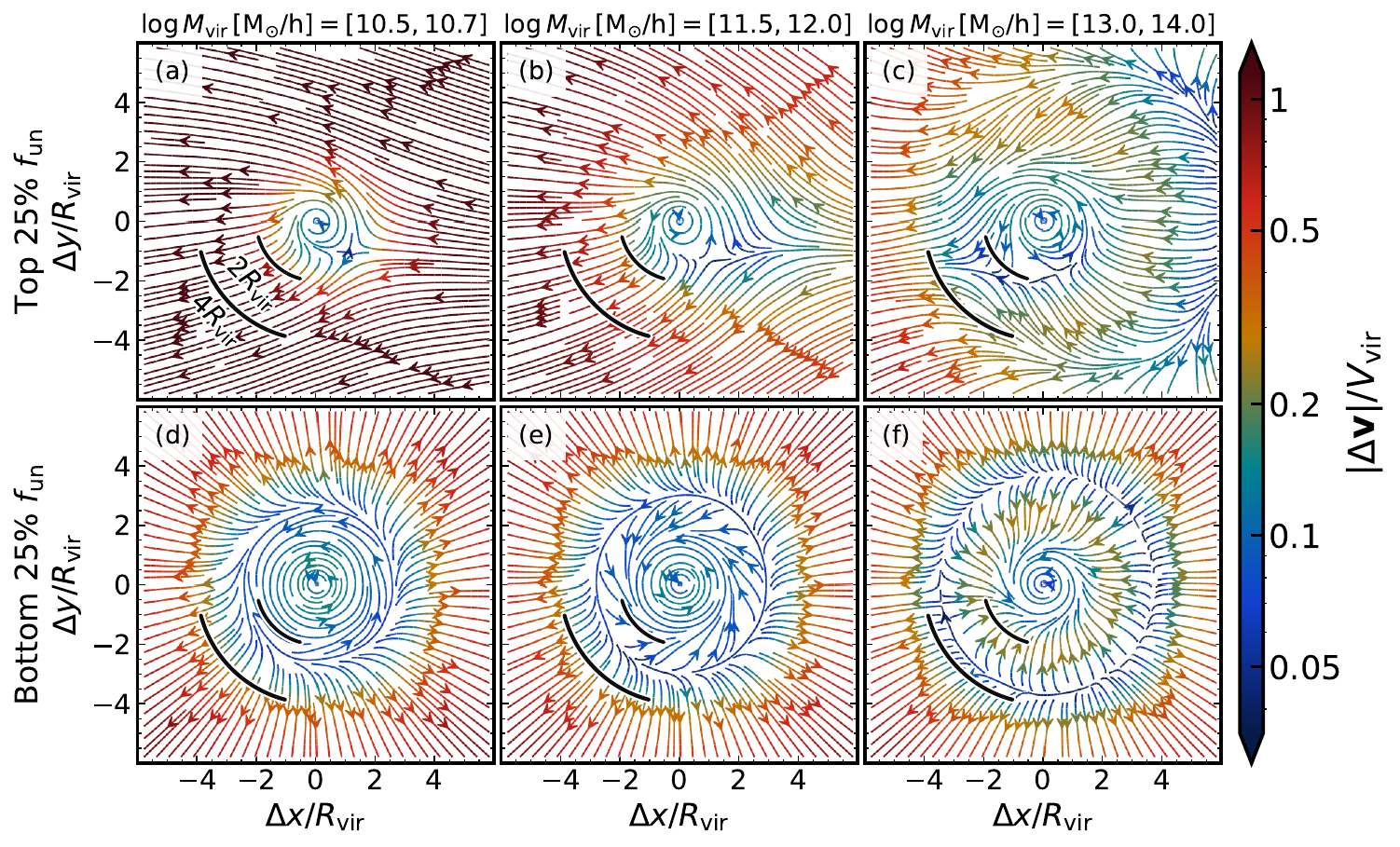 }
    \caption{
    {\figem Stacked velocity fields around haloes.}
    {\figem (a)}--{\figem (c)}, stacked velocity fields around haloes in 
    the upper quartile of $f_{\rm un}$ in each bin of $M_{\rm vir}$.
    {\figem (d)}--{\figem (f)}, similar but for haloes in the lower quartile 
    of $f_{\rm un}$. Here the field around each halo is the physical velocity 
    (peculiar plus Hubble flow) in the rest frame of the halo 
    (referred to as the halo-centric frame). The velocities are measured
    within $\pm 0.25\,R_{\rm vir}$ of the $\Delta x$--$\Delta y$ plane
    and projected onto that plane.
    The stacking follows the method developed by \citet{maFountainPatternBaryon2026}, 
    with the halo-centric frame aligned so that 
    the velocity vectors of the haloes point in the $+\Delta x$ direction,
    and the spin vectors of the haloes are closest to the $+\Delta z$ 
    direction. In each panel,
    the arrows of the streamlines show the direction and the colour encodes 
    the magnitude of stacked velocity (normalized by the average virial 
    velocity of haloes being stacked);
    the two black arcs indicate the halo-centric distances 
    of $2R_{\rm vir}$ and $4R_{\rm vir}$, respectively.
    This figure shows that haloes with different $f_{\rm un}$ exhibit flow patterns 
    with distinct divergence and vorticity in their circum-halo medium 
    (CHM), indicating that environment regulates halo accretion through
    modifying the flow pattern of the CHM (see \S\ref{ssec:fun-tracing-flow} 
    for details).
    }
    \label{fig:sample_large_scale_velocity_stream}
\end{figure*}

\begin{figure*}
    \includegraphics[width=0.96\textwidth]{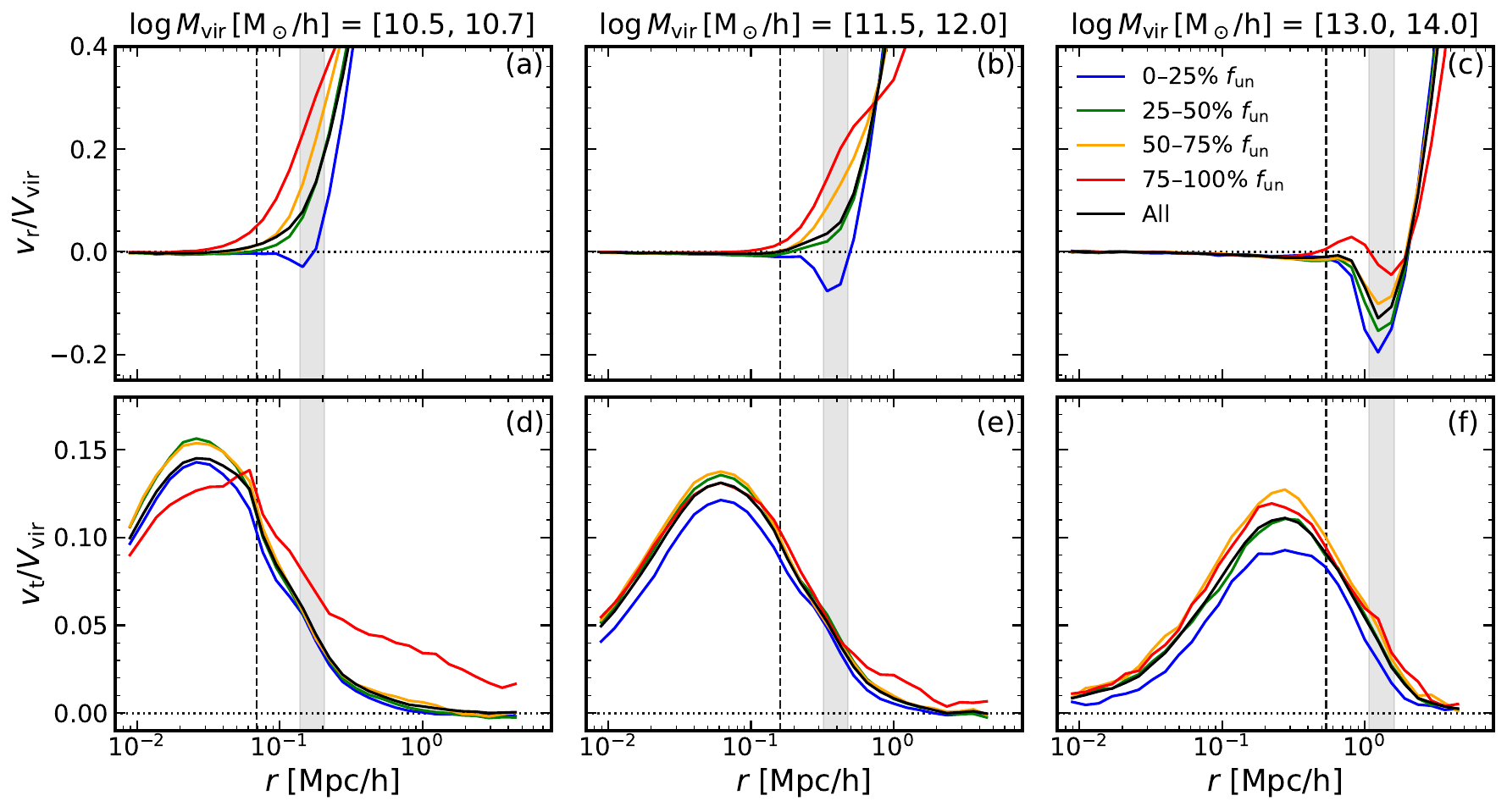}
    \caption{
    {\figem Stacked radial and tangential velocity profiles of haloes.}
    Here we show the physical velocity in the rest frame of the halo,
    stacked in spherical shells at different halo-centric distances
    around haloes in each bin of $M_{\rm vir}$ and $f_{\rm un}$.
    Two components of the velocity, the radial component $v_{\rm r}$ 
    (panels {\figem a--c}; projected onto the direction from halo 
    centre to the particle, with a positive value indicating outflow)
    and the tangential component $v_{\rm t}$
    (panels {\figem d--f}, projected along the direction perpendicular
    to both the radial direction and the spin axis of halo),
    are shown separately.
    The figure is obtained in a similar way to Fig.~\ref{fig:sample_large_scale_velocity_stream}, 
    using the stacking method of \citet{maFountainPatternBaryon2026} 
    and the same alignment strategy.
    In each panel, the vertical dashed line marks the average virial radius 
    $R_{\rm vir}$ of haloes being stacked, 
    while the vertical shaded region indicates the radial range of 
    $2$--$3\,R_{\rm vir}$ where $f_{\rm un}$ is measured.
    This figure quantifies the flow patterns of CHM revealed in 
    Fig.~\ref{fig:sample_large_scale_velocity_stream},
    with $v_{\rm r}$ and $v_{\rm t}$ serving as proxies for the (integrated)
    divergence and vorticity of the flow, respectively (see 
    \S\ref{ssec:fun-tracing-flow} for details).
    }
    \label{fig:sample_large_scale_velocity_radial}
\end{figure*}

\begin{figure} \centering
    \includegraphics[width=\columnwidth]{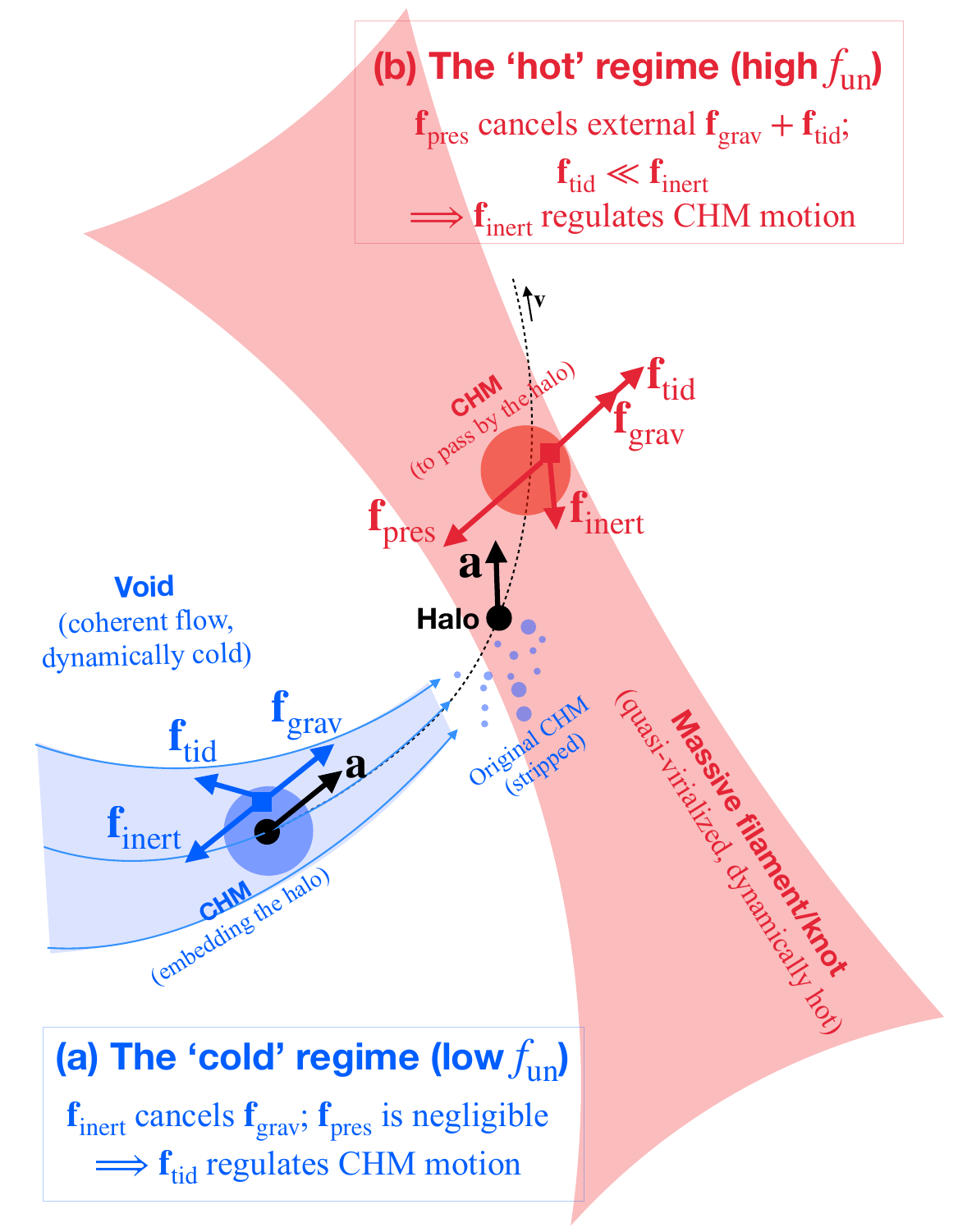}
    \caption{
    {\figem A schematic illustration of the two regimes of secondary bias.}
    The two regimes, the `cold' (numbered {\figem a}; corresponding to 
    low $f_{\rm un}$ and small halo bias) and `hot' (numbered {\figem b}; corresponding 
    to high $f_{\rm un}$ and large halo bias) 
    regimes, are characterized by the distinct dynamical hotness of the 
    large-scale environment in which the target halo resides, producing
    different force cancellation, leaving distinct forces that regulate the 
    flow pattern of the circum-halo medium (CHM) and the accretion onto 
    the halo.
    Secondary bias in the cold and hot regimes is thus driven by different 
    mechanisms, and a halo evolved from the former to the latter 
    regime may have experienced a switch in the dominant mechanism.
    Notations: $\bm{f}_{\rm grav}$, the uniform part of the gravitational force;
    $\bm{f}_{\rm tid}$, the tidal (i.e. differential) part of the gravitational 
    force; $\bm{f}_{\rm inert}$, the inertial force in the halo's rest frame,
    produced by the acceleration ($\bm{a}$) of the halo;
    $\bm{f}_{\rm pres}$, the pressure force from the random motion in a partially 
    virialized structure.
    See \S\ref{sssec:two-regimes} for details.
    }
    \label{fig:AssemblyBias-regimes}
\end{figure}

\section{Physical processes traced by the unbound matter}
\label{sec:physics-of-fun-vs-b}

The above correlation analyses imply a mediating role of $f_{\rm un}$ in linking
the environments and haloes, but have not yet revealed the underlying physical
mechanisms through which the unbound matter operates on haloes. These analyses,
however, do provide directions for searching for the relevant physical processes:
the significance of $\alpha$ in controlling secondary bias indicates that 
the accretion flow can be modified by the presence of large-scale structure; 
at smaller scales, interactions between haloes, with splashback events 
as the extreme cases, are known to have direct and diverse impacts, such as 
truncating the accretion or stripping material.
These processes point to a common scenario, in which the environment modifies the 
flow pattern of the {\bfem circum-halo medium} (CHM), which in turn affects the 
accretion rate and geometry, and hence the internal properties of haloes; 
CHM accelerated to a high velocity relative to the motion of a halo is likely unbound from 
the halo, and is thus captured by $f_{\rm un}$.
In this section, we use stacked fields to show that $f_{\rm un}$ indeed traces 
the flow pattern of the CHM, and to infer the physical processes at large 
(\S\ref{ssec:fun-tracing-flow}) and small scale 
(\S\ref{ssec:fun-tracing-compet}) that drive the pattern.

\subsection{$f_{\rm un}$ tracing the divergence and vorticity of circum-halo flow}
\label{ssec:fun-tracing-flow}

% Discuss the pattern; minimize discussion of physics here
\subsubsection{The pattern of circum-halo flow}
\label{sssec:flow-pattern}

To gain an intuition for how the flow pattern of the CHM is related to $f_{\rm un}$,
in Fig.~\ref{fig:large_scale_velocity_stream}, we overlay the
streamlines of peculiar velocity on the density field (grey scales) of the slice 
containing the same two example haloes as those in Fig.~\ref{fig:fun_example}.
For clarity, we hereafter refer to the coordinate system of the simulation and to
that of the zoom-in views as the simulation frame and the halo-centric frame 
(i.e. the halo's rest frame, centred on the halo),
respectively, and denote the coordinates (velocities) in the two frames by
$\bm{x}$ ($\bm{v}$) and $\Delta \bm{x}$ ($\Delta \bm{v}$), respectively.
To be consistent with the definition of binding energy (\S\ref{ssec:halo_property}),
$\Delta \bm{v}$ includes the Hubble expansion.

Comparing the two cases in the two frames reveals striking differences:
(i) In the simulation frame (panel a), the low-$f_{\rm un}$ halo is embedded
in a coherent, large-scale ($\approx 5\mpc$), high-speed ($\approx 500\kms$)
bulk flow whose velocity is comparable to that of the halo.
The halo is therefore a `co-moving observer'
of the ambient medium, so that the relative velocity between the halo and the CHM
is small. This small relative velocity implies that the CHM is effectively
bound to the halo, which results in a low $f_{\rm un}$ in the CHM and
in a halo-centric flow pattern converging on the halo (panel c).
(ii) In contrast, the high-$f_{\rm un}$ halo is embedded in a convergent,
low-speed ($\approx 100\kms$), filamentary structure
in the simulation frame (panel a), while the halo itself moves
much faster than the flow ($\approx 500\kms$). The halo therefore behaves as a `subhalo-like
observer' within a partially virialized, dynamically hot structure,
mimicking the condition that would be experienced by a subhalo falling
into a well-virialized host halo.
The motion of the CHM relative to such a halo is set by the virial velocity of
the massive host structure, yielding a relative velocity field of large
amplitude.
Consequently, only a small fraction of the CHM can be bound to the halo,
the accretion onto the halo is suppressed, and the halo-centric frame shows a
unidirectional, non-convergent flow (panel b).

The two cases reveal distinct CHM flow patterns in both frames.
A low-$f_{\rm un}$ halo tends to be embedded in a
coherent (dynamically cold) flow in the simulation frame and to co-move with it; 
in contrast, a high-$f_{\rm un}$ 
halo tends to move randomly within a partially virialized (dynamically hot) 
structure, so that it behaves like a subhalo.
Transforming to the halo-centric frame, the flow therefore has
a distinct divergence (i.e. $\nabla \cdot \bm{v}$), converging onto
the low-$f_{\rm un}$ halo and divergently passing by the high-$f_{\rm un}$ one.

To test the representativeness of the pattern seen in these two cases, we take haloes from
each bin of $M_{\rm vir}$ and select subsamples of haloes whose $f_{\rm un}$
values are in the upper and lower quartiles within
the same $M_{\rm vir}$ bin. We then stack the halo-centric velocity fields
around each subsample, using the stacking method developed by
\citet{maFountainPatternBaryon2026}.
To reveal both the passing-by motion (identified above) and spinning motion (see below)
of CHM, the halo-centric frames are aligned so that the velocity vector of a halo points in the $+\Delta x$ direction, and the spin vector 
(computed from the same set of particles as used for $\lambda_{\rm a}$)
is closest to the $+\Delta z$ direction. 
%\todo[YC]{Check that this is indeed the alignment method in use}.
To provide complementary diagnostics of convergence and rotation, we also obtain 
the halo-centric profiles of the
radial velocity ($v_{\rm r}$, with a positive sign indicating outflow)
and of the tangential velocity ($v_{\rm t}$, perpendicular to $v_{\rm r}$ and to the
spin vector) by stacking the corresponding velocity component
of the CHM in spherical shells centred on the haloes in each subsample
(see \citealt{maFountainPatternBaryon2026}). 
In Figs.~\ref{fig:sample_large_scale_velocity_stream} and 
\ref{fig:sample_large_scale_velocity_radial}, we show the stacked 
velocity field in the $\Delta x$--$\Delta y$ plane, and
the velocity profiles, respectively, of the CHM for each subsample.

The stacked fields show patterns consistent with the two cases 
(Fig.~\ref{fig:large_scale_velocity_stream}b, c), and reveal more:
\begin{enumerate}
\item
The high-$f_{\rm un}$ subsamples
(Fig.~\ref{fig:sample_large_scale_velocity_stream}a--c)
all exhibit a passing-by flow out to $\gtrsim 2R_{\rm vir}$,
but a clear mass dependence
is visible: the more massive haloes (panel c) have a lower CHM velocity
(in the halo-centric frame, scaled by $V_{\rm vir}$), and the velocity direction
is more radial within $\approx 2$--$4\,R_{\rm vir}$.
The $v_{\rm r}$ profiles provide complementary evidence: 
the highest-$f_{\rm un}$ quartile has outward median motion outside 
$R_{\rm vir}$ in the two lower-mass bins (Fig.~\ref{fig:sample_large_scale_velocity_radial}a, b),
but retains an infall dip at high mass (panel c).
\item
The low-$f_{\rm un}$ subsamples
(Fig.~\ref{fig:sample_large_scale_velocity_stream}d--f) all exhibit a flow converging
onto the halo centres, but the range over which inflow dominates is mass
dependent. A circle-like ridge at which the radial velocity reverses sign is present around each subsample, and the radius of the ridge, $R_{\rm ta}$ (the turnaround radius; see \citealt{moGalaxyFormationEvolution2010}),
expands from $\approx 2R_{\rm vir}$ to $\approx 4R_{\rm vir}$
as $M_{\rm vir}$ increases from $10^{10.5}\msun$ to $10^{14}\msun$.
More massive haloes have larger inflow speeds within $R_{\rm ta}$,
and the inflow is more radially biased.
This mass dependence is also seen in the $v_{\rm r}$ profiles of
the lowest-$f_{\rm un}$ quartile: a negative-$v_{\rm r}$ minimum is
reached in the $2$--$3\,R_{\rm vir}$ shell, deepening with mass; 
the infall dip is confined to this lowest quartile in the two lower
mass bins, but is present in all quartiles in the highest mass bin
(Fig.~\ref{fig:sample_large_scale_velocity_radial}a--c).

\item
The CHM exhibits spinning motion in both the low- and high-$f_{\rm un}$
subsamples, but the effect is again mass dependent. 
In the lowest mass bin, the rotation is obvious in the low-$f_{\rm un}$ stack, but is superposed on a stronger passing-by flow
in the high-$f_{\rm un}$ stack (Fig.~\ref{fig:sample_large_scale_velocity_stream}a, d).
The $v_{\rm t}$ profiles resolve the radial dependence of this
contrast: the
highest-$f_{\rm un}$ quartile has larger $v_{\rm t}/V_{\rm vir}$ outside
$R_{\rm vir}$, indicating a higher vorticity
($\nabla \cross \bm{v}$) of the CHM,
yet a lower $v_{\rm t}/V_{\rm vir}$ within $R_{\rm vir}$ (Fig.~\ref{fig:sample_large_scale_velocity_radial}d)
than the lowest quartile. In the two higher mass bins, the
high-$f_{\rm un}$ population has a higher inner $v_{\rm t}/V_{\rm vir}$
than the lowest quartile.
\end{enumerate}

%%%% The physical mechanisms/processes, i.e. the driving 'forces' underlying the 
%%%% pattern; namely, linking the motion to the force via Newton's second law
\subsubsection{Two regimes of halo-environment connection}
\label{sssec:two-regimes}

The distinct flow patterns around the two populations with low and high
$f_{\rm un}$, and the preference of the populations to reside in distinct 
large-scale environments 
(indicated by the map in Fig.~\ref{fig:large_scale_velocity_stream}a, 
and the correlation of $f_{\rm un}$ with $\alpha$ and $b$
in \S\ref{sec:secondary_bias}), lead to a hypothesis 
that two regimes of secondary bias exist. In this hypothesis, the 
large scale environment (as characterized by $\alpha$ and $b$) determines 
the driving factors that regulate the flow pattern (including
the divergence and vorticity) of CHM, and the flow pattern in turn 
determines the accretion onto haloes and hence their internal properties.

% Dynamical equation (Newton's law) is the fundamental physics -- the 
% distinction in forces thus means fundamentally different nature.
To illustrate the two regimes, and to argue that they are physically distinct, 
we seek the forces that drive the CHM motion. We refer the reader to 
Fig.~\ref{fig:AssemblyBias-regimes} for a schematic outline. Consider a 
mass element in the CHM surrounding, or about to pass by a halo in question. 
In the halo-centric frame, this mass element is subject to a net force 
(per unit mass) of
\begin{equation}
    \bm{f}_{\rm chm} =
        \bm{f}_{\rm grav}
        + \bm{f}_{\rm tid}
        + \bm{f}_{\rm pres}
        + \bm{f}_{\rm inert}\,,
\end{equation}
where $\bm{f}_{\rm grav}$ is the gravitational force averaged over all mass 
elements in the CHM (i.e. the uniform part of gravity across the CHM);
$\bm{f}_{\rm tid}$ is the tidal force, describing the differential gravity
across the CHM of finite size; 
both $\bm{f}_{\rm grav}$ and $\bm{f}_{\rm tid}$ include the contribution 
from large-scale environment (i.e. the matter outside the CHM, plus the 
cosmological constant), and the halo itself; 
$\bm{f}_{\rm pres}$ is the `thermal pressure' force caused by random motion
when the CHM belongs to a structure with a certain degree of virialization;
and $\bm{f}_{\rm inert}$ is the inertial force, which is a virtual 
force defined to be the negative of the halo acceleration.
In the special case that a halo moves in an orbit around some centre, 
$\bm{f}_{\rm inert}$ reduces to the centrifugal force;
in general, the halo may not have a well-defined orbit 
(e.g. Fig.~\ref{fig:AssemblyBias-regimes}a, b).
%or if the halo is not %trapped
%in an orbit like that in %Fig.~\ref{fig:AssemblyBias-regimes}b),
%our formulation using $\bm{f}_{\rm inert}$ still applies.

%{\color{red} I think a better name for the inertial force is `ram pressure' force
%due to the streaming motion of the CHM relative to the halo.
%Energy conservation gives $v^2 (r_{\rm sub}) = GM_{\rm host}/r_{\rm sub}$. 
%Thus the ram pressure force per particle is 
%$${dv^2\over dr_{\rm sub}} \approx {GM_{\rm host}\over r_{\rm sub}^2}
%$$
%Energy is a state quantity, an integration of %the gravitational acceleration 
%over time.  The gravity term is the gravity of %the halo on CHM, namely 
%the binding force. $f_{\rm un}$ should be a %competition 
%between $f_{\rm grav}$ and all other three %forces (ram pressure, 
%thermal pressure, and tidal force.
%Cold means thermal pressure is small, comoving %means ram pressure is small. 
%Thus only tidal force operates in that case. %In hot flow, thermal pressure 
%is important, ram pressure is not important if %the random velocity is 
%higher than the relative velocity between the %halo and the average of 
%CHM particles.}   

We consider the following two extreme cases, as identified through their 
distinct $f_{\rm un}$, CHM flow pattern, and large-scale environment 
(\S\ref{sssec:flow-pattern}), and examine each term that contributes to 
$\bm{f}_{\rm chm}$:
\begin{enumerate}
    \item {\bfem The `cold' regime}, where low-$f_{\rm un}$ haloes co-move 
    with their CHM embedded in coherent flows at large scale (e.g. voids;
    Fig.~\ref{fig:AssemblyBias-regimes}a). 
    In this case, the CHM is attached to the halo throughout the halo 
    assembly history; the condition of coherent motion implies cold 
    dynamics (i.e. the bulk motion dominates over the random motion,
    in terms of the kinetic energy partition; see e.g. 
    \citealt{maTwophaseFormationGalaxies2026} for quantifiers), 
    so that the `pressure' force is negligible
    \begin{equation}
        \bm{f}_{\rm pres} \approx 0 \,;
    \end{equation}
    in addition, the `co-moving' motion implies that the halo and the CHM
    share the same acceleration of the coherent bulk flow, so that the
    uniform part of the gravitational field is cancelled by the inertial
    force in the halo-centric frame. These lead to
    \begin{equation}
        \bm{f}_{\rm grav} + \bm{f}_{\rm inert} \approx 0 \,.
    \end{equation}
    The vanishing or cancellation of these forces
    leaves the tidal (differential) gravity the only force
    that can drive the relative motion. 
    In this regime, whenever the halo's own gravity
    dominates over the external tide, the CHM is naturally gathered
    onto the halo, giving rise to the infall-dominated pattern
    (Fig.~\ref{fig:sample_large_scale_velocity_stream}d--f).
    {\textem The tidal field is therefore identified as the dominant factor 
    regulating the motion of CHM relative to the halo,
    and plausibly underlies the secondary bias of low-$f_{\rm un}$ haloes} 
    (see \S\ref{sssec:bias-interp} for details).

    \item {\bfem The `hot' regime}, where high-$f_{\rm un}$ haloes move 
    like subhaloes hosted within quasi-virialized, massive structures 
    (high $\alpha$ or $b$; e.g. massive filaments or the outskirts of nodes).
    In this regime, the CHM originally associated with 
    the halo has already been stripped away,
    and the CHM currently surrounding the halo, or to pass by the
    halo at some later epochs, is part of the virialized structure
    (Fig.~\ref{fig:AssemblyBias-regimes}b).
    Hence, the virialization condition implies hot dynamics in the host 
    structure, and that the pressure support due to the random motion 
    cancels the external gravity on the CHM from the large-scale 
    environment:
    \begin{equation}
        \bm{f}_{\rm grav}^{\rm (ext)} + \bm{f}_{\rm tid}^{\rm (ext)} 
        + \bm{f}_{\rm pres} \approx 0\,,
        \label{eq:hot-cancellation}
    \end{equation}
    leaving only the inertial force to operate on the CHM, and to compete 
    with the gravity of the halo itself when the CHM is passing by the halo.
    In the case that the host structure is not perfectly virialized, 
    the cancellation in Eq.~\ref{eq:hot-cancellation} may be imperfect.
    However, the sub-dominance of the external tide can still be argued 
    from the following order-of-magnitude estimate:
    assuming the host structure has a mass $M_{\rm host}$, the inertial 
    force generated on the CHM by the subhalo-like orbital motion of 
    the halo in question is
    \begin{equation}
        \left|\bm{f}_{\rm inert}\right| \approx 
        \frac{G M_{\rm host}}{r_{\rm sub}^2}\,,
    \end{equation}
    where $r_{\rm sub}$ is the distance of the halo to the centre of the host structure, and $G$ is the gravitational constant;
    the external tidal force on the CHM of a finite size $\delta l_{\rm chm}$ 
    at a distance $r_{\rm chm}$ from the host structure is 
    \begin{equation}
        \left|\bm{f}_{\rm tid}^{\rm (ext)}\right| 
        \lesssim 4\pi G\bar\rho \lambda_{1} \delta l_{\rm chm} 
        \approx \frac{G M_{\rm host}}{r_{\rm chm}^3} \delta l_{\rm chm}\,,
    \end{equation}
    where $\lambda_{1}$ is the largest eigenvalue of the tidal tensor
    (see Eq.~\ref{eq:tidal-tensor-def}).
    For CHM that is passing by, or is about to pass by, the halo,
    $r_{\rm chm} \sim r_{\rm sub}$, so that the ratio of their magnitudes is 
    \begin{equation}
        \frac{\left|\bm{f}_{\rm tid}^{\rm (ext)}\right|}{\left|\bm{f}_{\rm inert}\right|}
        \lesssim \frac{\delta l_{\rm chm}}{r_{\rm sub}}
        \sim \frac{R_{\rm vir}}{r_{\rm sub}}\,,
    \end{equation}
    where $\delta l_{\rm chm} \sim$ a few times $R_{\rm vir}$ by the 
    `circum-halo' definition.
    Thus, the subdominance of the external tidal force,
    $\left|\bm{f}_{\rm tid}^{\rm (ext)}\right| \ll \left|\bm{f}_{\rm inert}\right|$,
    holds for any halo whose size is much smaller than its distance to the 
    centre of the host structure. 
    Consequently, {\textem the inertial force that arises from the 
    assembly of massive, dynamically hot structures and the motion of 
    haloes within those structures, rather than the tidal field, is the most 
    plausible driver of the passing-by motion, and hence plausibly
    underlies the secondary bias of high-$f_{\rm un}$ haloes}
    (see \S\ref{sssec:bias-interp} for details).
\end{enumerate}

Although our force-based argument applies to the instantaneous acceleration 
of the CHM, the time-integrated effects of the forces can be compared
as well: the quasi-virialization of the host structure implies a conservative 
inertial force and consequently path-independent work;
the path integrals of the $\bm{f}_{\rm inert}$ and 
$\bm{f}_{\rm tid}^{\rm (ext)}$ can then be evaluated along a straight-line
infall orbit of the halo into the host structure, 
and the subdominance of the tide at every point along the path implies that 
the integrated work done by the tidal force (and thus the resulting relative velocity
between the CHM and the halo) is also subdominant compared to
that by the inertial force.

While the above two regimes do not represent the entire halo population, 
they bracket the flow pattern and the driving forces at the two extremes, and 
the general halo population is expected to be a mixture of the two cases. 
In an evolutionary sense, a CDM halo may start as a local density
peak embedded in a coherent flow (i.e. in the cold regime), 
and then gradually fall into a dynamically hot, massive structure 
(i.e. in the hot regime). During this evolution
the two driving forces, $\bm{f}_{\rm tid}$ and $\bm{f}_{\rm inert}$, 
compete or synergize with each
other in shaping the flow pattern around and the accretion onto the halo, 
and tend to dominate the secondary bias when the halo is in the 
corresponding regime.
The switching of the dominant factor of the secondary bias, 
and the accompanying increase in the dynamical hotness of the embedding
environment, are therefore {\textem a reflection 
of the hierarchical nature of structure formation in CDM cosmology},
while our proposed indicator, $f_{\rm un}$, is a sensitive 
tracer of this formation process.

The two regimes, and the evolutionary connection between them, also imply
a mass dependence of the dominant factor of the CHM flow.
A low-mass halo can start in a void and migrate to a massive 
filament or node, completing the switch from
the cold to the hot regime, and from $\bm{f}_{\rm tid}$ to
$\bm{f}_{\rm inert}$ as the dominant factor.
In contrast, a massive halo (e.g. Fig.~\ref{fig:sample_large_scale_velocity_stream}c), 
especially an extremely massive one, is unlikely to be embedded in a 
more massive structure, since these structures do not have enough time 
to assemble in the finite age of the Universe;
the migration of such a halo to the hot regime is therefore incomplete, 
and its situation is more similar to that in the cold regime.

\subsubsection{The two-regime interpretation of secondary bias}
\label{sssec:bias-interp}

The two-regime scenario not only offers an interpretation of the CHM flow patterns, 
but also provides a route to understanding the essential characteristics 
in the secondary bias of different internal halo properties.

For low-mass haloes, the inertial-force-driven passing-by flow in the hot 
regime prevents the CHM from being bound to, and delivered into the haloes,
as supported by their high $f_{\rm un}$ (Fig.~\ref{fig:diffm_un_bias}),  
and the divergence in stacked velocity fields 
(Fig.~\ref{fig:sample_large_scale_velocity_stream}a, d)
and $v_{\rm r}$ profiles 
(Fig.~\ref{fig:sample_large_scale_velocity_radial}a).
This reduces the accretion of the haloes in their later 
assembly histories, when they migrate around and into massive host structures, 
and leaves the material assembled at early times, when the haloes 
are still in the cold regime, the main contributor to halo mass and
other internal properties.
The stalled later accretion accounts for their larger $z_{1/2}$, the steeper 
density profile, the reduced outer boundaries 
(see Appendix~\ref{app:halo-boundary}),
and hence higher $c_{\rm vir}$, 
of low-mass haloes in denser environments (higher $b$ or 
$\alpha$; Fig.~\ref{fig:spearman_fun_delta}a, b).

The stalled accretion at later times also explains spin bias of low-mass haloes.
In a dense, highly anisotropic environment, high-vorticity motion 
is induced in the CHM (Figs.~\ref{fig:sample_large_scale_velocity_stream}a, 
\ref{fig:sample_large_scale_velocity_radial}d).
However, the inertial-force-driven passing-by flow in the hot regime
prevents the spinning material in the CHM from being bound to, and delivered into 
the halo, so that the halo retains only the low-spin material assembled at early 
times. In contrast, a halo in less dense environment develops less tangential 
motion in the CHM (Figs.~\ref{fig:sample_large_scale_velocity_stream}d, 
\ref{fig:sample_large_scale_velocity_radial}d), 
but its inward motion (i.e. more negative divergence) allows 
angular momentum to be delivered to the halo, thus producing higher
spin of bound material.
Such a scenario is consistent with the inverted rank of $v_{\rm t}$ profile 
within $R_{\rm vir}$ between the low- and high-$f_{\rm un}$ quartiles
(see Fig.~\ref{fig:sample_large_scale_velocity_radial}d),
and explains the opposite sign of the $\lambda_{\rm b}$-$b$ correlation 
to that of the $\lambda_{\rm a}$-$b$ correlation at low mass
(Fig.~\ref{fig:spearman_fun_delta}a).

The incomplete migration of massive haloes to the hot regime, as discussed in 
\S\ref{sssec:two-regimes}, limits the inertial-force-driven passing-by flow;
this, together with the stronger gravitational attraction of massive haloes,
allows the CHM to be bound to, and delivered into, the haloes.
Such a mass-dependent switch of regimes is consistent with the lower
$f_{\rm un}$ (Fig.~\ref{fig:fun_props}), the weaker flow divergence,
and the more radially biased motion
around more massive haloes (Figs.~\ref{fig:sample_large_scale_velocity_stream},
\ref{fig:sample_large_scale_velocity_radial}).
This also offers a plausible explanation for the fact that
CDM haloes exhibit a universal pattern of two-phase assembly
and that more massive haloes transition to the
slow-accretion phase later \citep[e.g.][]{zhaoGrowthStructureDark2003,
moTwophaseModelGalaxy2024}.

The capability of massive haloes to retain the CHM allows the angular momentum
to be delivered into the halo, and thus explains the inverted $\lambda_{\rm b}$ 
bias at intermediate mass and the increasingly positive $\lambda_{\rm b}$ bias 
at high mass
(Fig.~\ref{fig:spearman_fun_delta}a).
The mass dependences of the $z_{1/2}$ and $c_{\rm vir}$ biases, however,
are more complex, and we refer the reader to
Appendix~\ref{app:internal-properties-in-alpha-qe-plane} for a detailed
interpretation.

% Link to previous studies.

The distinction between the two regimes suggests why simple tidal-field
metrics, such as the tidal anisotropy ($\alpha$), the
classical Hill radius \citep[$R_{\rm Hill}$; see e.g. Eq.~10 of][]{hahnTidalEffectsEnvironment2009} and analogous formulations,
may not fully account for the secondary bias of haloes
\citep{mansfieldThreeCausesLowmass2020}, since the halo-centric frame
is not inertial, especially for the high-$f_{\rm un}$ haloes,
and the CHM acted on by the inertial force is not attached to the halo.
In particular, the subdominance of the tidal force in the high-$f_{\rm un}$ 
haloes implies that the reduction of residual secondary bias after 
controlling for tidal metrics, such as $\alpha$ (Fig.~\ref{fig:spearman_alpha_fun}) and 
$R_{\rm Hill}$ \citep{mansfieldThreeCausesLowmass2020},
does not necessarily indicate that the tidal field produces the
forces that regulate halo accretion;
instead, these metrics act as classifiers of the cosmic web: a
strongly anisotropic tidal field signals the formation of a massive,
dynamically hot, filamentary structure, whose motion generates the
inertial forces that suppress the accretion.
This agrees with \citet{Borzyszkowski2017ZOMGHowCosmic}, who show that for haloes
embedded within thick filaments, the infalling material joins the
filament rather than the halo, so that accretion onto these haloes
is suppressed; in our force formulation, it is the inertial force, and 
the passing-by motion of the CHM it induces, that is responsible for such 
suppression.
In Appendix~\ref{app:alpha-bias} and Fig.~\ref{fig:sample_large_scale_velocity_stream_alpha}, we show that the CHM flow 
pattern is indeed distinct between haloes with different $\alpha$, 
providing further support for the classifier role of the tidal metrics.

We note that our two-regime scenario provides a unifying framework,
as well as an evolutionary view, that connects the two scenarios
proposed in the literature to explain the secondary bias. 
The subdominance of tides in the hot regime
corresponds to the gravitational/dynamical `heating' scenario proposed 
and examined by e.g. \citet{moPreheatingPrevirializationIts2005}, 
\citet{wangEnvironmentalDependenceCold2007}, 
and \citet{mansfieldThreeCausesLowmass2020},
in which the random motion associated with the formation of large-scale 
structures is inferred to suppress the accretion onto haloes.
In particular, particles surrounding older, low-mass haloes are
found to have larger velocities relative to the halo
\citep[fig.~8 of][]{wangEnvironmentalDependenceCold2007},
linking the availability of slowly moving, capturable material to 
the formation-time bias.
Our flow pattern and force-based formulation clarify this scenario by
suggesting that it is the relative motion of the halo and the host
structure, not the random motion within the host structure
itself, that limits the accretion, and it is the inertial force that 
drives the relative motion.
In contrast, the dominance of tides in the low-$f_{\rm un}$ case corresponds to
the scenario of tide-limited accretion proposed and discussed by e.g.
\citet{wangInternalPropertiesEnvironments2011}, \citet{hahnTidalEffectsEnvironment2009},
and \citet{ramakrishnanCosmicWebAnisotropy2019}, in which the tidal shear
is found to correlate with halo properties.
Our flow pattern and force-based formulation clarify this
scenario by suggesting that such a driving role of tides is only valid 
in the cold regime.

% Thus, the formulation of Hill radius, r_Hill ~ d (m/(3M))^{1/3} (Eq. 10)
% in Hahn et al. 2009 cannot be applied to the high-f_un case -- the inertial 
% force here is the dominant factor.

% \citet{DiazAxtle2026InfluenceLargescaleStructure} similarly report
% that haloes with $M\gtrsim10^{12}\Msun$ are found almost exclusively
% in regions of low absolute divergence and low vorticity, while highly
% dynamical environments are dominated by low-mass haloes, consistent
% with the mass-dependent inflow pattern found here.

% The distinction between suppressed and sustained inflow is also
% reflected in the outer halo structure. As shown in
% Appendix~\ref{app:halo-boundary}, high-$f_{\rm un}$ subsamples have
% smaller splashback and depletion radii than their low-$f_{\rm un}$
% counterparts, particularly at low mass. These boundary differences
% complement the kinematic evidence for different halo growth states.

% The link to the cosmic web is further supported by
% \citet{sotosuarezTracingGalaxyBias2026}, who show that the proximity
% to massive nodes plays a central role in shaping galaxy bias.
% In other words, tidal diagnostics such as
% $\alpha$ account for both cases only when
% they serve as descriptors of the environment rather than as measures
% of the tidal force itself; a pure tidal-force characterisation is therefore insufficient. 

\begin{figure*}
    \includegraphics[width=0.9\textwidth]{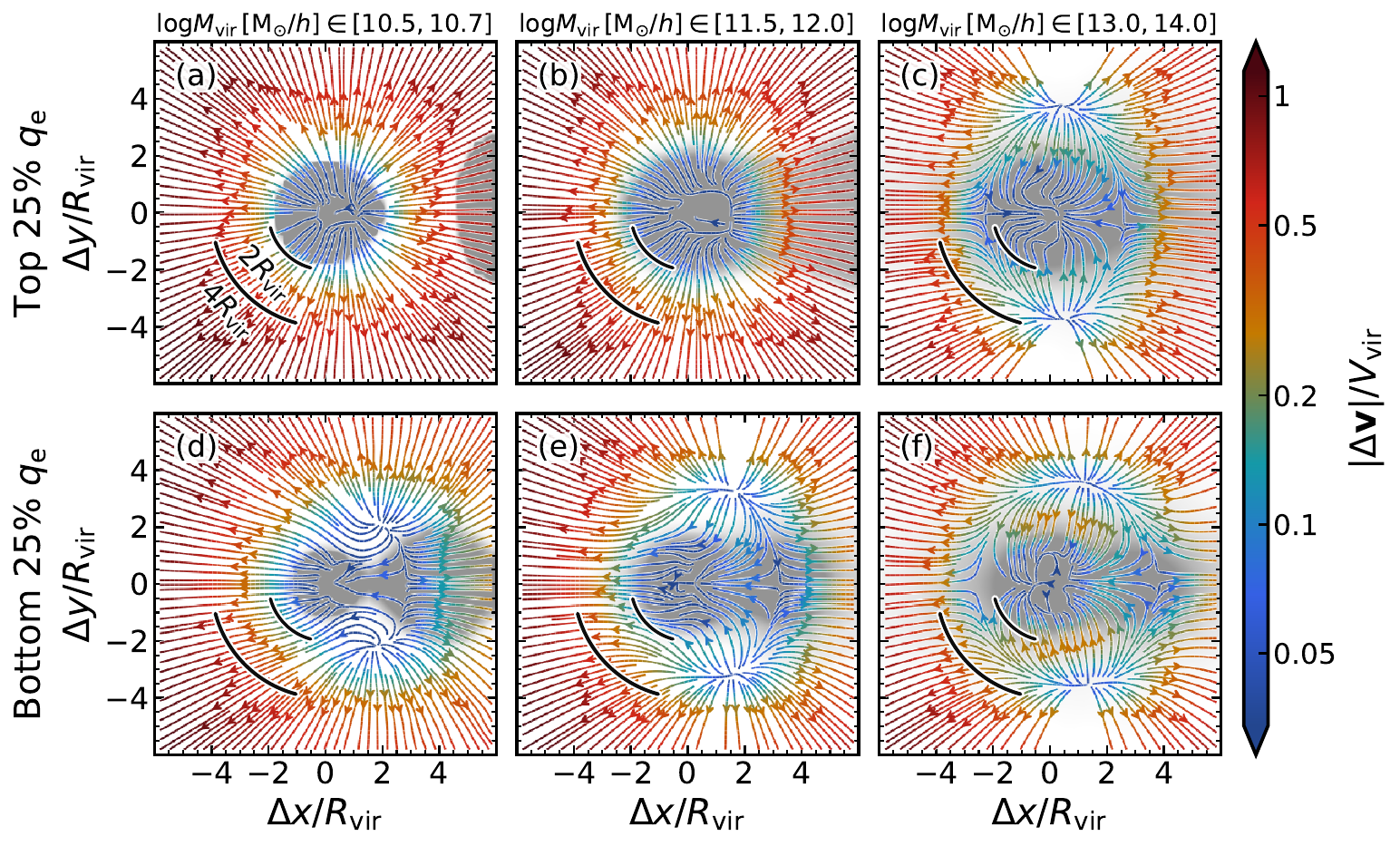}
    \caption{
    {\figem Stacked velocity fields around haloes with different environmental 
    axis ratio $q_{\rm e}$.}
    The figure is similar to Fig.~\ref{fig:sample_large_scale_velocity_stream}, 
    but the division of the halo sample is based on the axis ratio, $q_{\rm e}$,
    of the environment at halo-centric distances of $2$--$4\,R_{\rm vir}$.
    To control the effect of tidal anisotropy $\alpha$,
    stacking is performed by weighting the haloes
    so that the distributions of $\alpha$ for the haloes in the upper and lower 
    quartiles of $q_{\rm e}$ are aligned to the same distribution 
    as that of all haloes in the same mass bin.
    To reveal the companion structures,
    the alignment of the halo-centric frame is chosen so that the major and 
    intermediate axes of the inertia tensor involved in the computation of 
    $q_{\rm e}$ point to the $\Delta x$ and $\Delta y$ directions within the 
    page, with requirements that the denser side outside $R_{\rm vir}$ 
    points to the $+\Delta x$ direction and that the spin vector 
    using all particles within $R_{\rm vir}$ points out of the page to 
    break the remaining degrees of freedom of reflection.
    In each panel, the greyscale background shows the median
    surface mass density in the same slice.
    This figure shows that haloes with lower $q_{\rm e}$ at fixed $\alpha$
    tend to have a more prominent companion, which enhances the binding of the 
    CHM but also competes for the accreted material (see \S\ref{ssec:fun-tracing-compet} 
    for details).
    }
    \label{fig:merger_AB_stack_map}
\end{figure*}

\begin{figure*}
    \includegraphics[width=0.92\textwidth]{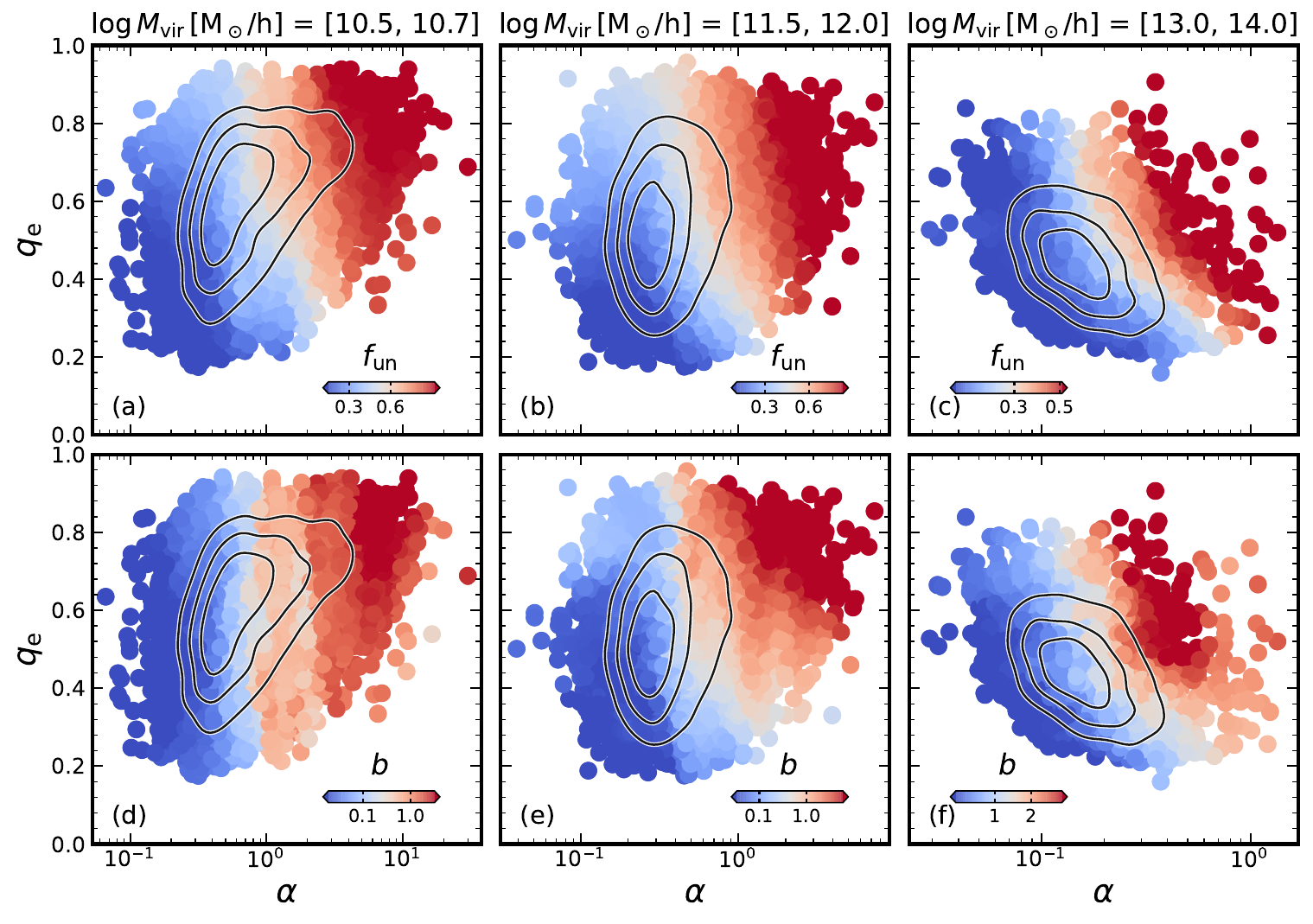}
    \caption{
    {\figem Unbound fraction $f_{\rm un}$ and bias $b$ as functions of tidal 
    anisotropy $\alpha$ and environmental axis ratio $q_{\rm e}$.}
    {\figem (a)}--{\figem (c)}, distribution of individual haloes (dots) 
    within bins of $M_{\rm vir}$ in the plane of $\alpha$ and $q_{\rm e}$, 
    coloured by $f_{\rm un}$.
    {\figem (d)}--{\figem (f)}, similar but coloured by $b$.
    In each panel, black contours encompass $25$, $50$ and $75$\% of the 
    haloes.
    The {\software Loess} algorithm \citep{clevelandRobustLocallyWeighted1979,Cappellari2013} is used to smooth the coloured property to reveal 
    the trend. This figure demonstrates that $b$ depends on environmental metrics
    at different scales, and that $f_{\rm un}$ traces $b$ in the combined space 
    of $\alpha$ and $q_{\rm e}$ (see \S\ref{ssec:fun-tracing-compet} for 
    details).
    }
    \label{fig:Correlation_alpha_Q_fun}
\end{figure*}

\subsection{$f_{\rm un}$ tracing competitive accretion}
\label{ssec:fun-tracing-compet}

The flow patterns in \S\ref{ssec:fun-tracing-flow} show how the
large-scale environment regulates the motion and boundedness of the CHM.
On smaller scales, a close companion can also modify the material
available to a halo. 
The splashback population in \S\ref{ssec:splashback}
illustrates such an effect in the presence of a past encounter with a more massive
system. 
Here we examine the current configuration of the surrounding matter,
particularly when two haloes compete for a common supply. 

To characterize the local geometry that cannot be fully specified by 
the smoothed tidal anisotropy, $\alpha$, we use the environmental axis ratio,
$q_{\rm e}\equiv a_3/a_1$, defined in \S\ref{ssec:halo_property} from all
particles in the shell with a halo-centric radius of $2$--$4\,R_{\rm vir}$.
A prominent companion halo increases the extent of the mass distribution 
along its major axis, producing a small $q_{\rm e}$, 
while a more nearly spherical distribution has a larger $q_{\rm e}$. 
Thus, $q_{\rm e}$ can distinguish the local configuration of 
matter at a given tidal environment at large scale.
% $q_{\rm e}$ distinguishes local configurations at fixed
% large-scale tide; since companions also contribute to the local
% potential, this geometry can affect both $f_{\rm un}$ and the
% accretion direction.

Fig.~\ref{fig:merger_AB_stack_map} compares the ambient density and
velocity fields of haloes in the upper and lower quartiles of
$q_{\rm e}$. Within each mass bin, the two samples are reweighted to a
common $\alpha$ distribution, so their differences cannot be attributed
simply to different distributions of tidal anisotropy. In the
low-$q_{\rm e}$ stacks (panels d--f), a prominent off-centre density
concentration appears along the positive major axis, whose sign is
chosen to point towards the denser side of the environment. The
high-$q_{\rm e}$ stacks (panels a--c) lack such a close companion,
but show a larger structure at greater distance.
At the low-mass end, the characteristic configuration is a halo beside a 
comparable or more massive neighbour, while at the high-mass end, the target 
halo is the more massive member of the pair. The high-mass, 
low-$q_{\rm e}$ stack also
shows an elongated distribution extending to the opposite side of the
halo, indicating that the pair is embedded in a larger, filamentary 
structure.

A companion affects the surrounding material in two ways: together
with the matter embedding the pair, it deepens the local potential and
increases the fraction of particles classified as bound, lowering
$f_{\rm un}$ in our energy estimate (\S\ref{ssec:halo_property}); at
the same time, it attracts part of the common supply and limits the
region from which the target halo can accrete, particularly
restrictive for a low-mass halo beside a more massive neighbour,
consistent with the tidal suppression of halo growth
\citep{hahnTidalEffectsEnvironment2009}.
A low $f_{\rm un}$ in such a configuration therefore indicates
substantial binding, but does not imply that all of the CHM will
contribute to the future growth of the target halo.
The relative importance of these effects varies with halo mass: a
massive target halo retains a larger share of the shared material,
with the deepened potential accompanied by inflow towards its own
centre (Fig.~\ref{fig:merger_AB_stack_map}c, f), whereas the flow
around a low-mass halo beside a companion is more asymmetric, more
restricted in space, and slower. 
This yields a scenario of `competitive accretion' at small scales, 
and predicts a stronger reduction of $f_{\rm un}$ at low 
$q_{\rm e}$ for high-mass haloes.

Fig.~\ref{fig:Correlation_alpha_Q_fun} allows us to examine this
expectation while separating the effects of $\alpha$ and $q_{\rm e}$.
The population contours show a positive $\alpha$--$q_{\rm e}$ relation
at low mass, a weaker dependence at intermediate mass, and a negative
relation at high mass. A possible geometrical explanation is that a
low-mass halo can be embedded within a much larger anisotropic
structure, while its relatively small CHM shell samples a locally
rounder distribution. Such a configuration is consistent with the
dynamically hot, passing-by CHM described in
\S\ref{ssec:fun-tracing-flow}. For a massive halo, the CHM extends over
scales comparable to the surrounding structure and resolves its
elongation, so a larger $\alpha$ can coincide with a smaller
$q_{\rm e}$. 

The colour gradients in Fig.~\ref{fig:Correlation_alpha_Q_fun}(a)--(c)
show that $f_{\rm un}$ generally increases with $\alpha$ within each mass bin. 
At a given $\alpha$, $f_{\rm un}$ also increases with $q_{\rm e}$, 
consistent with the deepening of the local potential in the presence of a 
companion.
The stronger conditional correlation between $f_{\rm un}$ and $q_{\rm e}$
at high mass is consistent with the scenario of
competitive accretion, 
reflecting the greater ability of more massive haloes to
retain the shared material. 

The colour patterns of halo bias $b$ in Fig.~\ref{fig:Correlation_alpha_Q_fun}(d)--(f) 
are similar to those of $f_{\rm un}$, and the density maps in 
Fig.~\ref{fig:merger_AB_stack_map} offer a plausible
explanation for this additional dependence of $b$ on $q_{\rm e}$:
the high-$q_{\rm e}$ stack has a rounder distribution within $\lesssim 4\,R_{\rm vir}$,
and lacks a prominent companion at this scale; however, it exhibits an
extended overdensity at $\gtrsim 4\,R_{\rm vir}$, indicating the presence of a
massive, larger-scale structure beyond the CHM, and yielding a higher bias
than the low-$q_{\rm e}$ counterpart at fixed $\alpha$.
The similarity between the dependence of $b$ and $f_{\rm un}$ on $q_{\rm e}$
at fixed $\alpha$ can also be understood: in order to generate a similar $\alpha$,
the surrounding matter can either be a massive, large-scale overdensity,
which yields a large $b$ at large scales but also a large $f_{\rm un}$ 
due to the lack of a close companion, or a less massive
companion, which lowers $f_{\rm un}$ within the CHM but reduces $b$
at large scales.

Our interpretation suggests that the configuration of matter on different
scales affects the fraction of unbound material, $f_{\rm un}$, and thus
accretion onto a halo, with $\alpha$ a major
proxy for that configuration and $q_{\rm e}$ a complementary one.
Indeed, \citet{mansfieldThreeCausesLowmass2020} found that a
single-neighbour tidal proxy does not account for all low-mass assembly
bias, and that tides from the extended matter distribution and the
velocities of ambient particles must also be considered. 
Therefore, the combination of $\alpha$ and $q_{\rm e}$ is expected to
account for much of the secondary bias, in a manner similar to
$f_{\rm un}$, as seen from the similar gradients of $f_{\rm un}$ and $b$ in 
the $\alpha$--$q_{\rm e}$ plane (Fig.~\ref{fig:Correlation_alpha_Q_fun}).

% This decomposition leads to a definite expectation: conditioning jointly on $\alpha$ and $Q$ should approach the effectiveness of conditioning on $f_{\rm un}$. 
% To test this expectation, we condition simultaneously on $\alpha$ and $Q$, using the same method as in Eq.~\ref{eq:conditional_alpha_Q}; the resulting residual secondary bias is shown as the grey dashed error bars in Fig.~\ref{fig:spearman_alpha_fun}. For the dynamical properties, namely shape, spin, and velocity anisotropy, the residuals after controlling for $\alpha$ and $Q$ are nearly comparable to those after controlling for $f_{\rm un}$, consistent with the expectation that for these properties the information carried by $f_{\rm un}$ is largely the combination of the environmental information encoded in $\alpha$ and the coherent-accretion information encoded in $Q$.

\section{Summary and discussion}
\label{sec:summary}

In this paper, we have used the {\software TNG300-1-Dark} simulation
to investigate how the binding state of the circum-halo medium (CHM) is
related to halo secondary bias. Our sample consists of haloes with
$M_{\rm vir}=10^{10.5}$--$10^{14.5}\msun$ at $z=0$.
The unbound mass fraction, $f_{\rm un}$, measured in the $2$--$3\,R_{\rm vir}$ 
shell around each halo, has a strong correlation with halo bias,
making it a sensitive indicator of secondary bias.
Our main findings are as follows.

\begin{enumerate}
\item
Haloes with larger $f_{\rm un}$ are more strongly clustered at every
mass examined, with a correlation coefficient  
$\Gamma_{f_{\rm un}b}\simeq0.35$--$0.6$.
Conditioning on $f_{\rm un}$ reduces the correlation coefficients
between bias and internal properties of haloes (formation time,
concentration, spins using bound or all particles, shape, or 
velocity anisotropy) to a few percent.
The reduction is nearly complete for spins and velocity anisotropy.
Formation time, concentration, and shape only retain weak residual
biases (\S\ref{ssec:fun-bias}).

\item We examine other environmental indicators that have been proposed to 
explain the origin of secondary bias.
Among them, tidal anisotropy is strongly correlated with 
$f_{\rm un}$, and both substantially
reduce secondary bias; $f_{\rm un}$ leaves smaller residuals for spin
and velocity anisotropy, while the relative performance 
depends on halo property and mass.
Low-mass splashback haloes have elevated $f_{\rm un}$, and conditioning
on $f_{\rm un}$ reduces secondary bias within both the splashback and
non-splashback samples. Weak residual biases in properties related to assembly
history remain (\S\ref{ssec:related-to-previous-mechanism}).

\item
The stacked velocity field and radial profiles reveal two distinct flow
patterns for the CHM around haloes with the lowest and highest $f_{\rm un}$:
one in which haloes
co-move with their CHM, and the other in which haloes move through a
dynamically hot, partially virialized environment. As a result,
the CHM around low-$f_{\rm un}$ haloes shows converging motion 
(more negative $\nabla \cdot \bm{v}$, smaller $\nabla \cross \bm{v}$),
while that around high-$f_{\rm un}$ haloes exhibits stronger passing-by
flow (less negative or positive $\nabla \cdot \bm{v}$, larger $\nabla \cross \bm{v}$). 
The flow pattern not only determines the ability of the halo to
accrete from the CHM, but also affects the delivery of angular momentum 
through the accretion.
The distinction depends on halo mass and is most pronounced at low
mass. At all masses, larger $f_{\rm un}$ accompanies reduced inflow.

\item
The physical mechanism producing secondary bias is thus inferred to be two-fold, 
with distinct driving forces in the two regimes: 
in the `cold' regime, tide-regulated CHM motion operates for 
low-$f_{\rm un}$ haloes that co-move with their CHM embedded in 
dynamically cold structures at large scales;
in the `hot' regime, tidal force is subdominant, and the inertial force 
generated by acceleration of haloes operates for high-$f_{\rm un}$ haloes
moving through massive, dynamically hot structure. 
As a halo evolves, the hierarchical nature of structure formation in CDM 
cosmology implies that the dominant mechanism can switch from the former to 
the latter (\S\ref{ssec:fun-tracing-flow}).

\item
The local configuration of the CHM traces nearby companions at
fixed tidal anisotropy, and the resulting competitive accretion.
At fixed $\alpha$, $f_{\rm un}$ increases
with the environmental axis ratio $q_{\rm e}$ measured over
$2$--$4\,R_{\rm vir}$, most strongly at high mass. Stacks matched in
$\alpha$ associate small $q_{\rm e}$ with elongated configurations
containing a prominent companion or filamentary structure. The
population-level $\alpha$--$q_{\rm e}$ relation changes from positive
at low mass to negative at high mass. A companion can deepen the
potential and lower $f_{\rm un}$ while competing for the share of the 
common supply to the target halo (\S\ref{ssec:fun-tracing-compet}).
\end{enumerate}

Our results suggest a physical picture with two regimes, at the 
extremes of a continuous spectrum of halo--environment interactions, 
in which inflow is regulated by tidal or inertial forces. 
Our statistical results support $f_{\rm un}$ as a common mediator
that traces both types of interaction.

Our picture also offers a physical interpretation of the adopted
$f_{\rm un}$ aperture. The $2$--$3\,R_{\rm vir}$ shell brackets
two characteristic boundary scales (Appendix~\ref{app:halo-boundary}): the splashback
radius, the first apocentre of recently accreted material, is located
at $\approx 1$--$2\,R_{\rm vir}$ for typical haloes \citep[e.g.][]{diemerDependenceOuterDensity2014,moreSplashbackRadiusPhysical2015}, while the
depletion radius, out to which halo growth has exhausted the
surrounding matter, typically lies at $\sim2.5\,R_{\rm vir}$
\citep[e.g.][]{Fong2021NaturalBoundary,hanManyBoundariesStratified2026}.
The aperture therefore samples the transition zone just outside the
splashback shell but within the depleted region, where the material
is most sensitive to the balance between the halo potential and
environment, and may thus best encode the halo--environment
interaction.

We caution that small residual secondary bias conditioned on $f_{\rm un}$
does not in general prove conditional independence, and matching 
present-day $f_{\rm un}$ does not erase differences in past 
accretion or encounters. The non-monotonic trends of concentration at 
high mass (Appendix~\ref{app:internal-properties-in-alpha-qe-plane})
further caution against assigning a single structure or growth history 
to each unbound fraction. 
Moreover, the force-based discussion of \S\ref{sssec:two-regimes} 
remains qualitative. It assumes that haloes co-move with the CHM
in the cold regime and that the host structure is quasi-virialized
in the hot regime.
Another limitation of our analysis and interpretation is that
the stacked flow does not determine which member of a close pair
ultimately accretes a particle (\S\ref{ssec:fun-tracing-compet}),
since binding in halo outskirts is sensitive to the surrounding potential 
and tidal field \citep[see appendix~D of][]{mansfieldThreeCausesLowmass2020}.
Following haloes and particles in the CHM and host structures 
across snapshots would test whether changes in $f_{\rm un}$ precede 
changes in accretion rate and geometry, verify the assumptions behind 
the two-regime scenario, and distinguish the material captured by a halo 
from that captured by its neighbours.

Definitions of energy and radial aperture also affect the 
interpretation of $f_{\rm un}$.
We have verified that extending the potential aperture to $5\,R_{\rm vir}$
does not change the main results. Nevertheless, the potential includes
surrounding structures, and its reference radius is finite, so that
$f_{\rm un}$ is an operational measure of present-day binding,
rather than a count of particles destined to escape to infinity
\citep[e.g.][]{behrooziUnboundParticlesDark2013,
mansfieldThreeCausesLowmass2020}. Similarly, median radial velocities
diagnose typical inflow or outflow, not a mass accretion rate. Tests
with alternative potential conventions, numerical resolutions,
redshifts, and cosmologies are needed before treating the preferred
shell or the measured correlations as universal.

The implications for galaxies require a further modelling effort.
If reduced dark-matter inflow accompanies reduced gas accretion,
galaxies in high-$f_{\rm un}$ haloes could exhaust their gas reservoirs
earlier. However, cooling, feedback, and interactions with the ambient
gas may alter both the flow patterns and the transfer of angular
momentum. These processes may be particularly violent for low-mass,
high-$f_{\rm un}$ haloes, around which the passing-by CHM has velocities
comparable to or greater than $V_{\rm vir}$ (Fig.~\ref{fig:sample_large_scale_velocity_stream}a),
and transonic or supersonic motion may arise for gas near the virial
temperature.
In this case, turbulent mixing between halo gas and the passing-by
CHM may promote entrainment of ambient gas rather than the removal of
halo gas \citep[e.g.][]{duFormationCompactElliptical2019,liShockinducedStrippingSatellite2023}. Such entrainment could transfer angular momentum
from the CHM to the halo gas and potentially lead to larger galaxy
sizes, providing a possible physical connection to the strong clustering
of diffuse galaxies
\citep{zhangUnexpectedClusteringPattern2025,MaInterpretingStrongClustering2026}.
A decoupling of gas and dark-matter spin has been reported on galactic
scales in simulations \citep[e.g.][]{velliscigAlignmentShapeDark2015,jiangDarkmatterHaloSpin2019,Yanggalaxysizehalospin2023}.
The flow patterns found here suggest that such a decoupling may also
arise on CHM scales, particularly around low-mass,
high-$f_{\rm un}$ haloes.  
Matched haloes in dark-matter-only and
hydrodynamical simulations could test this possibility and establish
whether $f_{\rm un}$ predicts gas supply, star formation, or galaxy
size.

Observational tests will require calibrated proxies for $f_{\rm un}$. 
Tidal anisotropy has been estimated for SDSS groups on a fixed scale of 
$\sim2\mpc$ \citep{Paranjape2018DependenceGalaxy}, providing information on the 
halo environment. Complementary kinematic information may be obtained from 
neighbouring galaxies, whose coherent motions have been used to construct a 
halo-spin proxy \citep[e.g.][]{Kim2025TowardObservational}. Such measurements 
suggest a possible route to probing circum-halo motions and, after calibration, 
the binding state traced by $f_{\rm un}$. The outer density profile offers 
a complementary diagnostic, given its dependence on $f_{\rm un}$ 
(Appendix~\ref{app:halo-boundary}). Combining these environmental, kinematic, 
and density measurements in mock catalogues would test whether they 
can trace $f_{\rm un}$ and recover its relation to clustering at fixed halo 
mass, accounting for smoothing scale, projection, and sample selection. 

Another approach is to incorporate $f_{\rm un}$ directly into models of 
the galaxy--halo connection. It could serve as a secondary variable in a 
decorated halo occupation model \citep[e.g.][]{hearinIntroducingDecoratedHODs2016}. 
Alternatively, an effective halo mass could be constructed from $M_{\rm vir}$ and 
$f_{\rm un}$, extending the approach of \citet{xuDependenceHaloBias2018}. 
This would turn the environmental dependence identified here into a testable 
model of galaxy clustering.

%%%%%%%%%%%%%%%%%%%% REFERENCES %%%%%%%%%%%%%%%%%%

% The best way to enter references is to use BibTeX:

\section*{Data Availability}

{\software TNG300-1-Dark} is available at 
\url{https://www.tng-project.org}. 

\section*{Acknowledgements}

Qinglin Ma thanks Cheng Li, Dandan Xu, Yifeng Zhou, Joop Schaye, Fangzhou Jiang 
and Jiaxin Han, and YC thanks Zhaozhou Li,
for helpful discussions. 
YC is supported by Double First-Class Discipline Construction-Astronomy 
Discipline at Nanjing University, the National Natural Science Foundation 
of China (Grant no. 12503014), and the Fundamental Research Funds for the 
Central Universities (Grant no. KG202502).
The computations and presentations in this paper are supported by various 
software: {\softwarestyle Scipy} 
\citep{virtanenSciPy10Fundamental2020}, {\softwarestyle Numpy} 
\citep{VanDerWaltNumPyArrayStructure2011}, {\softwarestyle Matplotlib} 
\citep{hunterMatplotlib2DGraphics2007}, {\softwarestyle Ipython} 
\citep{perezIPythonSystemInteractive2007} and {\softwarestyle Astropy} 
\citep{robitailleAstropyCommunityPython2013,
astropycollaborationAstropyProjectBuilding2018,
astropycollaborationAstropyProjectSustaining2022}.
The authors acknowledge the use of ChatGPT (OpenAI), Claude (Anthropic) and 
DeepSeek as a writing and research aid in the preparation of this manuscript,
including literature searches, drafting assistance, and language editing; 
all scientific content, analysis, and conclusions are the authors' own.

\bibliographystyle{mnras}
\bibliography{references} % if your bibtex file is called example.bib

\appendix

\section{The selection of the radial range of unbound particles}
\label{app:radial-range-of-unbound-particles}

\begin{figure*}
    \includegraphics[width=0.95\textwidth]{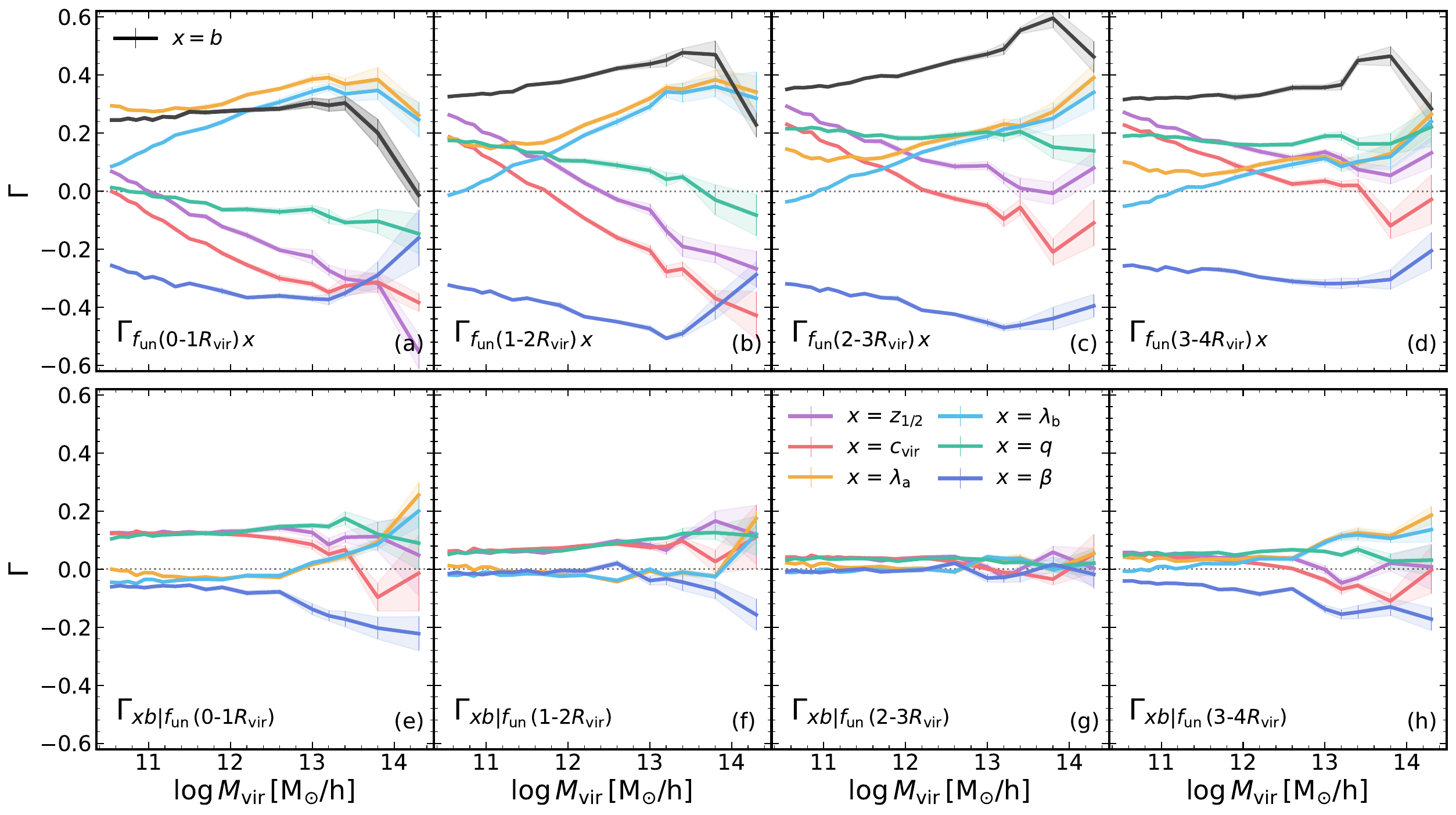}
    \caption{
    {\figem Spearman correlation coefficients using $f_{\rm un}$ measured in different 
    radial ranges.} 
    Each column shows the results similar to Fig.~\ref{fig:spearman_fun_delta}
    but using $f_{\rm un}$ measured in a given radial 
    range. The upper row shows the correlations of $f_{\rm un}$ with
    halo bias $b$ (black) and the internal properties (coloured).
    The lower row shows the partial correlations between $b$ and the
    internal properties after conditioning on $f_{\rm un}$.
    }
    \label{fig:spearman_fun_delta_diffr}
\end{figure*}

In the main text, our default choice of the radial range to measure
$f_{\rm un}$ is $2$--$3\,R_{\rm vir}$, where its correlation with 
halo bias is strongest (Fig.~\ref{fig:diffm_un_bias}c). Here we justify 
this choice by repeating the correlation analyses of 
\S\ref{ssec:fun-bias} for three alternative radial ranges, 
$(0$--$1)$, $(1$--$2)$, and $(3$--$4)\,R_{\rm vir}$. In Fig.~\ref{fig:spearman_fun_delta_diffr}, we show the Spearman correlation 
coefficients $\Gamma_{f_{\rm un} x}$ between $f_{\rm un}$ and each 
internal halo property $x$, along with $\Gamma_{f_{\rm un} b}$
(panels a--d), and the corresponding
conditional correlation coefficients $\Gamma_{xb \mid f_{\rm un}}$ by 
controlling for $f_{\rm un}$ (panels e--h), with each column 
adopting a different radial range to evaluate $f_{\rm un}$. 

The innermost shell, $(0$--$1)\,R_{\rm vir}$, behaves differently
from the others (Fig.~\ref{fig:spearman_fun_delta_diffr}a, e).
Within the virial radius, the unbound particles are rare
($f_{\rm un} \lesssim 0.008$; see Fig.~\ref{fig:fun_props}), and
including them in the halo definition directly biases the
measurements of the internal properties themselves: the inferred
concentration is depressed because the unbound particles
preferentially reside in the outer regions \citep[see
also][]{MaAnisotropicSecondaryBias2026}, while the spin is enhanced
and the velocity anisotropy reduced, consistent with their larger
tangential velocities and angular momenta. In this aperture,
therefore, $f_{\rm un}$ largely traces the contamination of the
internal-property measurements rather than the halo environment, and
its ability to account for the secondary bias is limited
(Fig.~\ref{fig:spearman_fun_delta_diffr}e).

For the outer shells, the correlations revert to the environmental
pattern seen in the main text: in low-mass haloes, a higher
$f_{\rm un}$ is associated with a higher concentration and an earlier
formation time (Fig.~\ref{fig:spearman_fun_delta_diffr}b--d). Among
the four apertures, the $(2$--$3)\,R_{\rm vir}$ shell yields the
strongest correlation with halo bias and, after conditioning on
$f_{\rm un}$, the smallest residual secondary bias
(Fig.~\ref{fig:spearman_fun_delta_diffr}c, g), consistent with the
pronounced separation of the $b$--$f_{\rm un}$ relations in
Fig.~\ref{fig:diffm_un_bias}(c). The $(1$--$2)\,R_{\rm vir}$ shell
gives intermediate results, while the $(3$--$4)\,R_{\rm vir}$ shell
remains correlated with halo bias but leaves larger residuals for
some properties (Fig.~\ref{fig:spearman_fun_delta_diffr}d, h).

We conclude that the choice of $2$--$3\,R_{\rm vir}$ is empirically optimal
within the radial ranges examined here. A plausible physical origin
of this aperture is discussed in \S\ref{sec:summary}: the shell
brackets the transition zone between the splashback and depletion
radii, where the boundedness of the surrounding material may be most
sensitive to the interaction between the halo and its environment.

\section{The role of tidal anisotropy in secondary bias}
\label{app:alpha-bias}

\begin{figure}
    \includegraphics[width=0.45\textwidth]{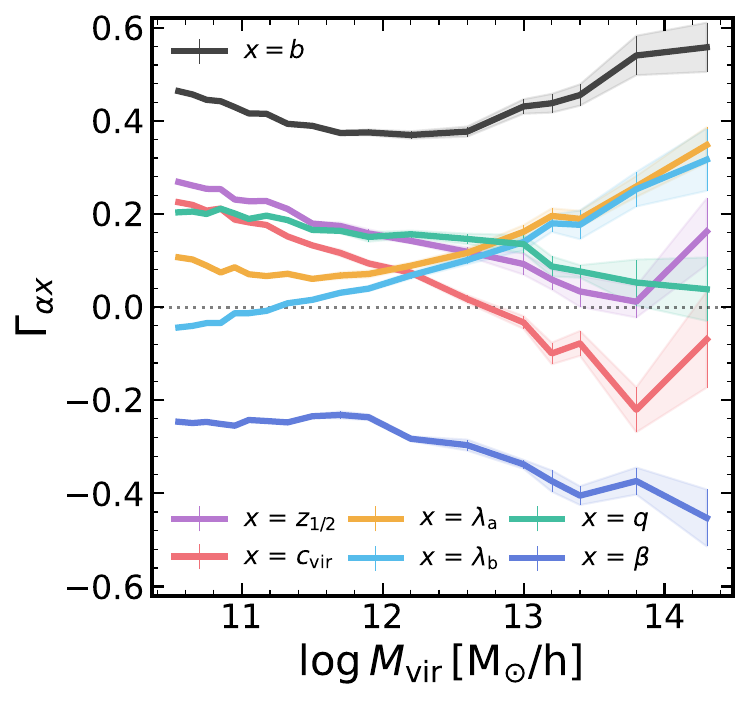}
    \caption{
    {\figem Spearman correlation coefficients between tidal 
    anisotropy $\alpha$ and halo properties.}
    Each coloured curve shows the correlation between $\alpha$ and an internal 
    halo property $x$ as a function of halo mass.
    For reference, the correlation between $\alpha$ and bias $b$ is shown as 
    the black curve.
    }
    \label{fig:spearman_alpha_delta}
\end{figure}

\begin{figure*}
    \includegraphics[width=0.98\textwidth]{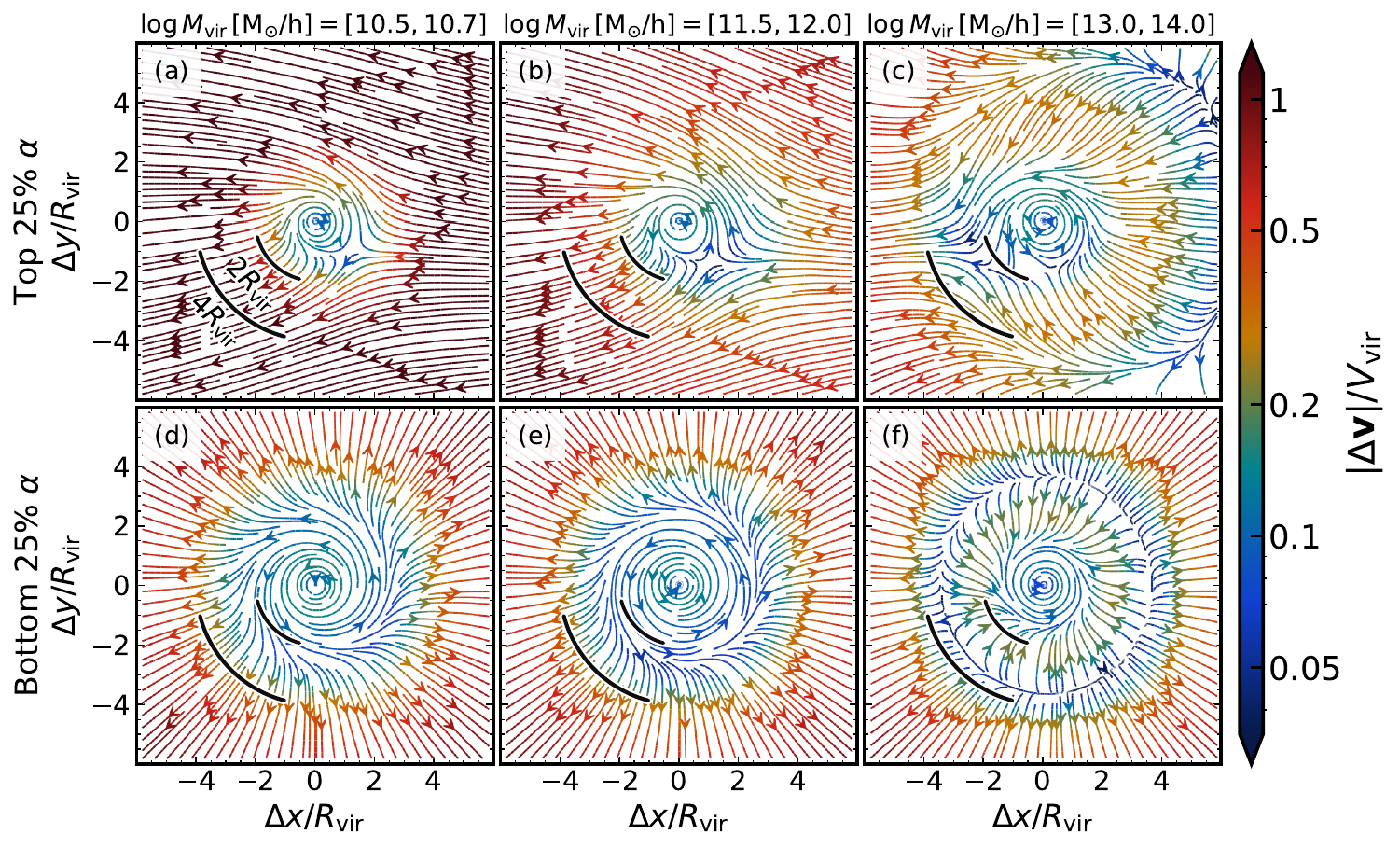}
    \caption{
    {\figem Stacked velocity fields of haloes split by tidal anisotropy $\alpha$.}
    As in Fig.~\ref{fig:sample_large_scale_velocity_stream}, the panels show the 
    velocity fields around haloes split by $\alpha$ in bins of halo mass.
    }
    \label{fig:sample_large_scale_velocity_stream_alpha}
\end{figure*}

In \S\ref{ssec:tidal-anisotropy}, we showed that the tidal anisotropy
$\alpha$ (see \S\ref{ssec:halo_property} for the definition) is strongly
correlated with $f_{\rm un}$, and that controlling for $\alpha$
suppresses the secondary bias. Here we complement the correlation
analysis by examining the correlation between $\alpha$ and the other 
halo properties, and supplement the flow analysis of 
\S\ref{ssec:fun-tracing-flow} by examining the CHM flows of haloes split by 
$\alpha$.

In Fig.~\ref{fig:spearman_alpha_delta}, we show the Spearman
correlation coefficients $\Gamma_{\alpha x}$ between $\alpha$ and 
each internal halo property $x$. Consistent with the expectation (i) of
\S\ref{ssec:tidal-anisotropy}, the correlations are significant over
the entire mass range examined, and their signs and mass dependences
closely mirror those of $\Gamma_{f_{\rm un} x}$
(Fig.~\ref{fig:spearman_fun_delta}b): haloes with higher $\alpha$ are
older, more concentrated, and rounder at low mass, and exhibit more
tangentially biased orbits. The conditional correlations
$\Gamma_{xb \mid \alpha}$ after conditioning on $\alpha$ are shown in
Fig.~\ref{fig:spearman_alpha_fun} of the main text, where conditioning
on $\alpha$ is found to largely suppress the secondary bias, in
agreement with \citet{ramakrishnanCosmicWebAnisotropy2019}.

Quantitative discrepancies remain, however. First, we measure a
stronger correlation between $\alpha$ and the bias factor $b$ across
the entire mass range (black curve in Fig.~\ref{fig:spearman_alpha_delta}), 
as well as stronger correlations between the secondary properties and $b$
(Fig.~\ref{fig:spearman_fun_delta}a), than reported by
\citet{Ramakrishnan2020SeparateUniversecalibration}. This difference
likely stems from the bias estimator: we measure the halo bias on
intermediate scales ($5$--$12\,h^{-1}\mathrm{Mpc}$), whereas
\citet{Ramakrishnan2020SeparateUniversecalibration} adopt a
power-spectrum method probing larger scales
($>30\,h^{-1}\mathrm{Mpc}$); measuring the bias on such large scales
may partially dilute the intrinsic correlation between $\alpha$ and
halo bias. Second, our most notable discrepancy lies in $\beta$: both
the $\beta$--$b$ and $\beta$--$\alpha$ relations exhibit a pronounced
mass dependence, whereas \citet{ramakrishnanCosmicWebAnisotropy2019}
observe almost no mass dependence. This discrepancy is likely
attributable to the definition of $\beta$: we consider only bound
particles within $R_{\rm vir}$, whereas
\citet{Ramakrishnan2020SeparateUniversecalibration} include bound
particles inside the mass ellipsoid 
%\todo[YC]{Check the beginning
%of their \S2.3, which says that only bound are used}. 
Consequently, $\alpha$ constrains
the secondary bias slightly differently in the two studies.
Nevertheless, conditioning on $\alpha$ substantially reduces
secondary bias.

In \S\ref{ssec:fun-tracing-flow}, we showed that the two flow patterns
of the circum-halo medium are traced by $f_{\rm un}$. Here we show
that the same dichotomy emerges when haloes are split by the tidal
anisotropy $\alpha$.
Fig.~\ref{fig:sample_large_scale_velocity_stream_alpha} presents the
stacked velocity fields constructed in the same way as
Fig.~\ref{fig:sample_large_scale_velocity_stream}, but with haloes
split within each mass bin by $\alpha$: the upper and lower rows show
haloes in the top and bottom 25\% of $\alpha$
(Fig.~\ref{fig:sample_large_scale_velocity_stream_alpha}a--c and
d--f, respectively).
The stacks support the expectation of
\S\ref{ssec:fun-tracing-flow}: haloes in highly anisotropic
environments have typically entered the vicinity of dynamically hot,
massive structures, and the resulting inertial force dominates the
motion of their CHM, driving the passing-by flow that elevates 
$f_{\rm un}$; in contrast, haloes in more isotropic environments co-move
with their CHM, and the tidal force governs their relative motion,
shaping converging, infall-dominated flows and lowering $f_{\rm un}$.
This agreement between the $\alpha$-split and $f_{\rm un}$-split flows
is consistent with the strong correlation between the two quantities found in
\S\ref{ssec:tidal-anisotropy}, and supports the interpretation that the 
tide is an indicator of the cosmic-web environment, rather than a direct 
driver of the CHM motion.

\section{The imprint of unbound mass fraction on the halo boundary}
\label{app:halo-boundary}

\begin{figure*}
    \includegraphics[width=0.98\textwidth]{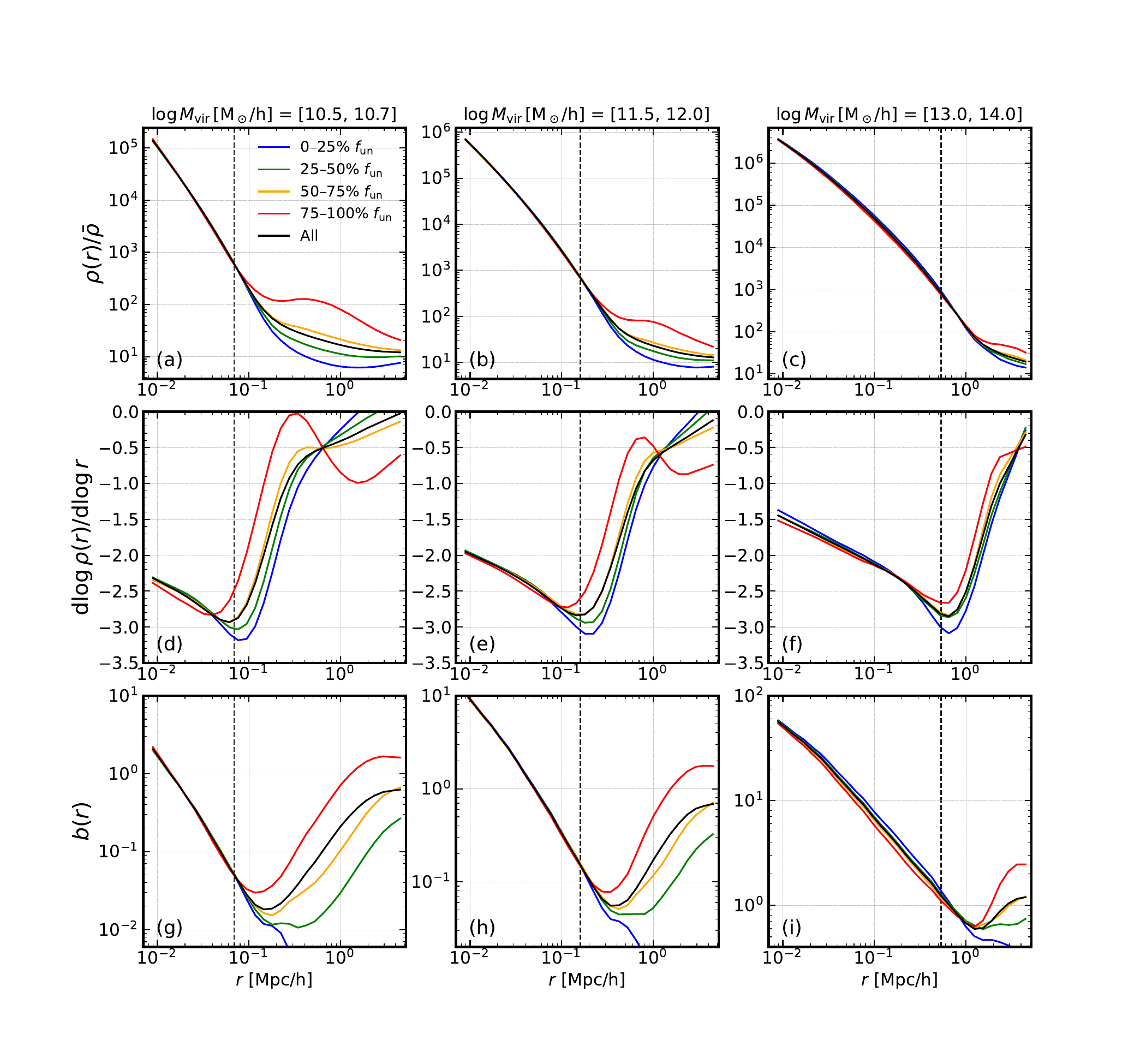}
    \caption{
    {\figem Radial profiles of density metrics around haloes}.
    The density metrics shown are the density contrast 
    $\rho(r)/\bar\rho$ (panels {\figem a}--{\figem c}),
    its logarithmic slope $\dd\ln\rho/\dd\ln r$ 
    (panels {\figem d}--{\figem f}), 
    and the bias factor $b(r)$ (panels {\figem g}--{\figem i};
    see Eq.~\ref{eq:bias} for the definition).
    Each column shows the results for haloes in a bin of $M_{\rm vir}$.
    Profiles in panels {\figem (a)}--{\figem (c)} are median values
    among haloes in use, and in panels {\figem (d)}--{\figem (f)}
    are the logarithmic slopes of the corresponding median density profiles, 
    and in panels {\figem (g)}--{\figem (i)} are the average value of the bias factor.
    In each panel, coloured curves show those profiles for haloes in 
    different quartiles of $f_{\rm un}$, while black curve shows that for 
    all haloes, with a vertical dashed line 
    marking the median virial radius of these haloes.
    }
    \label{fig:sample_large_scale_density}
\end{figure*}

In \S\ref{ssec:fun-tracing-flow}, we showed that high-$f_{\rm un}$
haloes experience reduced inflow relative to low-$f_{\rm un}$ haloes
of the same mass. 
The growth state of a halo has been found to relate to the outer density 
profile and to the boundary radii: rapidly accreting haloes develop steep outer
profiles whose logarithmic-slope minimum marks the splashback radius,
the first apocentre of recently accreted material
\citep[e.g.][]{diemerDependenceOuterDensity2014,moreSplashbackRadiusPhysical2015,shiOuterProfileDark2016},
while the growth of the halo depletes the surrounding matter out to a
depletion radius of $\sim2.5\,R_{\rm vir}$, revealed as a minimum of
the halo--matter bias profile
\citep[e.g.][]{Fong2021NaturalBoundary,hanManyBoundariesStratified2026}. 
It is thus expected that high- and low-$f_{\rm
un}$ haloes should exhibit systematically different boundary
structures. 
To test this, in Fig.~\ref{fig:sample_large_scale_density}, we show the
density profile $\rho(r)/\bar\rho$ (a--c), its logarithmic slope
(d--f), and the bias profile $b(r)$ (g--i; see Eq.~\ref{eq:bias} for 
the definition), for haloes in the same mass bins and $f_{\rm un}$ quartiles 
as in Fig.~\ref{fig:sample_large_scale_velocity_radial}. 

At low mass, the high-$f_{\rm un}$ haloes exhibit a higher concentration, 
although the difference is subtle in this representation
(Fig.~\ref{fig:sample_large_scale_density}a--c). More evident is that
high-$f_{\rm un}$ haloes show an enhanced density around the halo
boundary, consistent with the ambient material associated with massive
host structures streaming past the haloes; moreover, both the splashback and 
the depletion radii, marked by the minima 
of the logarithmic slope and of the bias profile, respectively
(Fig.~\ref{fig:sample_large_scale_density}d--f, g--i), are located at
smaller radii. Low-$f_{\rm un}$ haloes, which are still gaining mass,
show the opposite behaviour: ongoing accretion enlarges the region
over which the halo shapes its surroundings, pushing both boundary
radii outward, consistent with \citet{diemerDependenceOuterDensity2014}.
This boundary difference provides a potential observational test of the 
$f_{\rm un}$--bias connection, since the outer density profile can be measured 
through weak lensing or galaxy counts.

\section{Interpreting secondary bias across halo mass}
\label{app:internal-properties-in-alpha-qe-plane}

\begin{figure*}
    \includegraphics[width=0.85\textwidth]{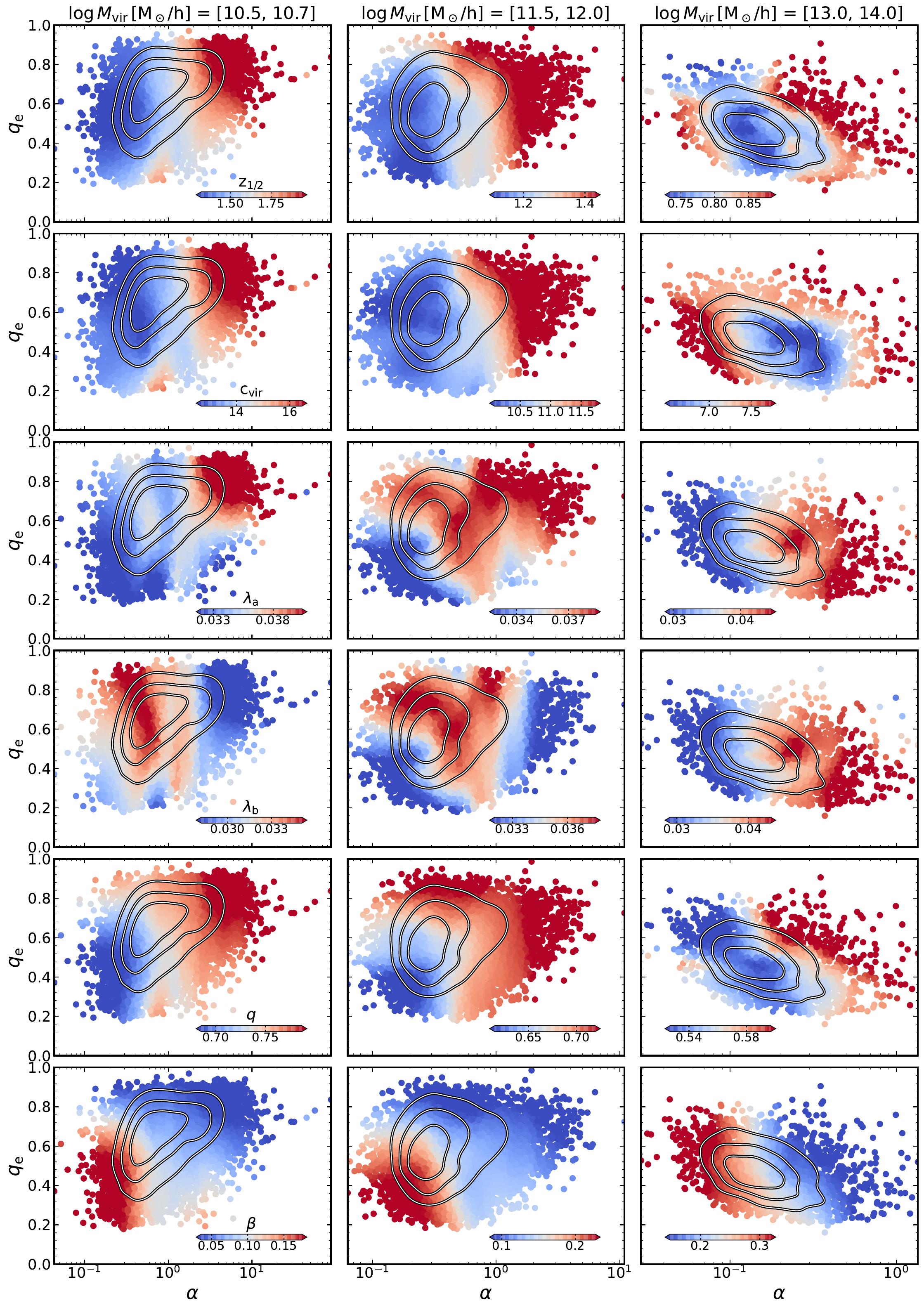}
    \caption{
    {\figem Internal properties in the $\alpha$--$q_{\rm e}$ plane.}
    The panels show individual haloes, colour-coded by formation time ($z_{1/2}$), concentration ($c_{\rm vir}$), spins in two definitions ($\lambda_{\rm a}$; $\lambda_{\rm b}$), shape ($q$) and velocity anisotropy ($\beta$) (from top to bottom), in three mass bins (from left to right).
    The contours encompass $25$, $50$ and $75$\% of the haloes in the $\alpha$--$q_{\rm e}$ plane.
    }
    \label{fig:concentration_alpha_fun}
\end{figure*}

In \S\ref{ssec:fun-bias}, we showed that the secondary bias of the
various internal properties differs in strength and even in sign. 
This appendix examines how the six internal properties 
(see \S\ref{ssec:halo_property} for the definitions) are related to
the two complementary environmental metrics, $\alpha$ and $q_{\rm e}$ 
(Fig.~\ref{fig:concentration_alpha_fun}), and uses 
these relations to interpret the characteristics of the secondary bias
of different internal halo properties across halo mass.

The six properties fall into three categories:
\begin{enumerate}
    \item The assembly metrics: formation time ($z_{1/2}$) and
    concentration ($c_{\rm vir}$). At low mass, both $z_{1/2}$
    and $c_{\rm vir}$ increase with $\alpha$ and $f_{\rm un}$. 
    Their gradients are similar to those of halo bias $b$ and $f_{\rm un}$ 
    in the $\alpha$--$q_{\rm e}$ plane (Fig.~\ref{fig:Correlation_alpha_Q_fun}a, d),
    consistent with the increased suppression of late accretion
    towards the `hot' regime or the lack of close neighbours 
    to deepen the binding potential, both preferring high-$b$ environments
    at large scales (\S\ref{sec:physics-of-fun-vs-b}).
    At high mass, however, a new population 
    emerges at the lower-left corner in the $\alpha$--$q_{\rm e}$ plane,
    in which both $z_{1/2}$ and $c_{\rm vir}$ are high 
    but $b$ is low (see Fig.~\ref{fig:Correlation_alpha_Q_fun}f).
    A possible origin for this population is that a massive halo without an 
    embedded overdensity (such as a filament or node)  
    may have exhausted its surroundings and stopped accreting,
    as indeed seen in the low bias profile and large depletion radius of 
    the high-mass, low-$f_{\rm un}$ haloes 
    (Fig.~\ref{fig:sample_large_scale_density}).
    A selection of high $z_{1/2}$ haloes thus includes those in both
    the upper-right (high $b$) and lower-left (low $b$) corners of the 
    $\alpha$--$q_{\rm e}$ plane, leading to a cancellation of $b$
    and thus weakened secondary bias of $z_{1/2}$ at high mass.
    The lower-left population is more prominent in $c_{\rm vir}$ than 
    in $z_{1/2}$, indicating that the concentration is more sensitive to the 
    exhausted accretion; a direct consequence is that a selection of high 
    $c_{\rm vir}$ haloes includes more low-$b$ haloes, which
    negates the average $b$ and reverses the secondary bias of $c_{\rm vir}$ at 
    high mass.

    \item    
    Spins ($\lambda_{\rm a}$ and $\lambda_{\rm b}$), defined using all particles and bound particles, respectively. The consistent
    sign of the $\lambda_{\rm a}$--$b$ correlation across halo mass,
    and the inverted sign of the $\lambda_{\rm b}$--$b$ correlation at 
    the low masses (Fig.~\ref{fig:spearman_fun_delta}a), 
    as already discussed in \S\ref{ssec:fun-tracing-flow},
    arise from a combination of two factors that emerge in the CHM pattern
    at high $f_{\rm un}$:
    the enhanced vorticity, which sources the angular momentum 
    of CHM, and the reduced negative divergence, which reduces the 
    delivery of angular momentum from CHM to the halo. 
    
    \item The anisotropy metrics: shape ($q$) and the velocity
    anisotropy ($\beta$). Both exhibit a monotonic gradient in the
    $\alpha$--$q_{\rm e}$ plane, aligning (or anti-aligning) 
    with the gradient of $f_{\rm un}$ 
    and $b$. The mass-invariant sign of the $q$--$b$ and
    $\beta$--$b$ correlation (Fig.~\ref{fig:spearman_fun_delta}a)
    indicates that these metrics are insensitive to the accretion
    rate, but rather reflect accretion geometry.
    The interpretation can also be made by our two-regime scenario 
    and competitive accretion picture: the passing-by flow in a `hotter' 
    (larger $\alpha$) 
    environment produces larger vorticity, and thus less radially 
    biased orbits (smaller $\beta$) and less elongation along 
    the accretion direction (larger $q$);
    the presence of a companion (lower $q_{\rm e}$) in contrast, enhances
    the accretion via competitive accretion, producing more 
    radially biased orbits (larger $\beta$) and more elongation along 
    the accretion direction (smaller $q$).
\end{enumerate}

\bsp	% typesetting comment
\label{lastpage}
\end{document}